\documentclass[epj]{svjour}
\usepackage{latexsym}
\usepackage{graphicx}
\usepackage{hyperref}
\usepackage{url}
\usepackage{amsfonts}
\usepackage{amsmath}
\usepackage{textcomp}
\usepackage{tikz}
\usepackage[sorting=none,maxnames=10,doi=false,url=false,date=year,isbn=false,eprint=false]{biblatex}
\newcommand\edm{8.3}
\renewbibmacro{in:}{}

\begin{document}
\title{Revisiting $^{129}$Xe electric dipole moment measurements applying a new global phase fitting approach}
\author{Tianhao Liu\inst{1,2,}\thanks{email:silasliutianhao@gmail.com}, Katharina Rolfs\inst{1,}\thanks{email:katharina.rolfs@ptb.de}, Isaac Fan\inst{1}, Sophia Haude\inst{1}, Wolfgang Kilian\inst{1}, Liyi Li\inst{2},Allard Schnabel\inst{1}, Jens Voigt\inst{1} \and Lutz Trahms\inst{1}}
\institute{ Physikalisch-Technische Bundesanstalt Berlin, 10587 Berlin, Germany \and Department of Electrical Engineering and Automation, Harbin Institute of Technology, 150001 Harbin, China}

\abstract{
By measuring the nuclear magnetic spin precession frequencies of polarized $^{129}$Xe and $^{3}$He, a new upper limit on the $^{129}$Xe atomic electric dipole moment (EDM) $ d_\mathrm{A} (^{129}\mathrm{Xe})$ was reported in Phys. Rev. Lett. 123, 143003 (2019). Here, we propose a new evaluation method based on global phase  fitting (GPF)  for  analyzing  the continuous phase development of the $^{3}$He-$^{129}$Xe comagnetometer signal. The Cramer-Rao Lower Bound on the $^{129}$Xe EDM for the GPF method is theoretically derived and shows the potential benefit of our new approach. The robustness of the GPF method is verified with Monte-Carlo studies. By optimizing the analysis parameters and adding data that could not be analyzed with the former method, we obtain a result of $d_\mathrm{A} (^{129}\mathrm{Xe}) = 1.1 \pm 3.6~\mathrm{(stat)} \pm 2.0~\mathrm{(syst)} \times 10^{-28}~ e~\mathrm{cm}$ in an unblinded analysis. For the systematic uncertainty analyses, we adopted all methods from the 
aforementioned PRL publication except the comagnetometer phase drift, which can be omitted using the GPF method. The updated null result can be interpreted as a new upper limit of $| d_\mathrm{A} (^{129}\mathrm{Xe}) | < \edm\ \times 10^{-28}~e~\mathrm{cm}$ at the 95\% C.L.}
\authorrunning{Liu et al.}
\maketitle

\section{Introduction}

A quantum field theory that models the formation of the imbalance of matter over antimatter in our universe must fulfill the Sakharov conditions \cite{Sakharov1967}. One of those conditions is the $\mathcal{CP}$ violation ($\mathcal{C}$ is charge conjugation and $\mathcal{P}$ is parity reversal). The best-tested standard model (SM) of particle physics provides two sources of $\mathcal{CP}$ violation, the phase of the Cabibbo-Kobayashi-Maskawa matrix and the   term $\bar{\theta}$ in the QCD Lagrangian \cite{Canetti2012}. However, the $\mathcal{CP}$ violation within the SM is too small to produce the observed rate of the matter to antimatter asymmetry, motivating searches for physics beyond-the-SM (BSM).  BSM theories generally include additional sources of $\mathcal{CP}$ violation \cite{Canetti2012,Dine2003}, such as a larger permanent electric dipole moment (EDM) of fundamental or composite particles \cite{Chupp2019,Cairncross2019}. So far, all measurement results of EDMs in more than ten diverse systems, with the first published in 1957 \cite{Smith1957}, are consistent with zero. These null results are interpreted as upper limits on EDMs and place constraints on various sources of $\mathcal{CP}$ violation and masses of BSM particles, thus directing the search of BSM scenarios \cite{Flambaum2020}. 

\begin{figure}[htpb]
    \centering{\includegraphics[width=0.6\columnwidth]{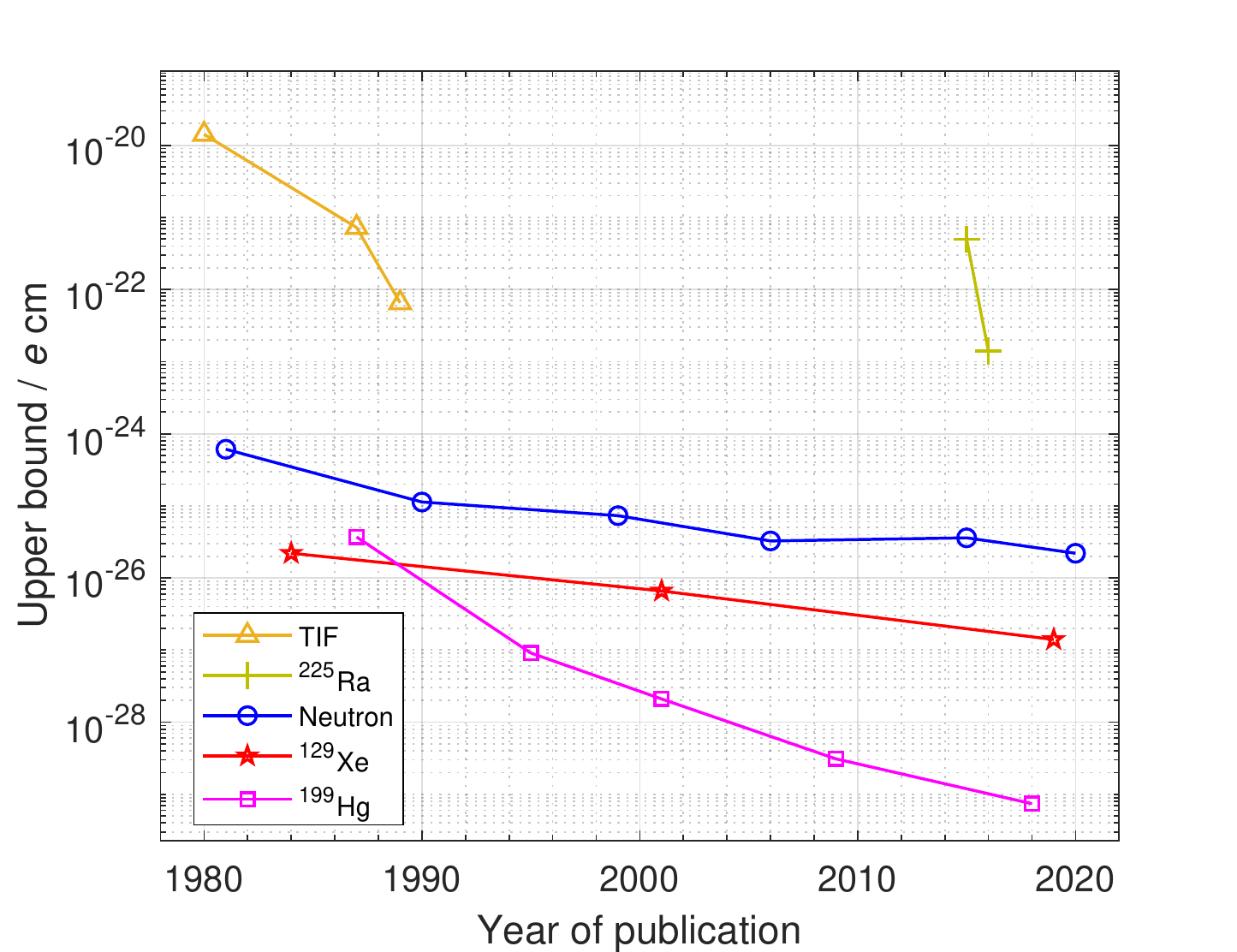}}
    \label{fig:EDM_all} 
    \caption{Selected upper bounds of EDM of diamagnetic systems performed since 1980 at the 95\% C.L. For all systems, the current upper bound has decreased more than an order of magnitude compared to their first published result \cite{Graner2016,Bishof2016,Sachdeva2019,Cho1989,Abel2020,Baker2006,Allmendinger2019,Rosenberry2001}.}
\end{figure}

Long spin-coherence time and obtainable high polarization leading to high signal-to-noise ratios (SNR) make several diamagnetic systems such as the $^{199}$Hg and $^{129}$Xe atom promising candidates for EDM experiments. Over the last 40 years, significant progress was made in the determination of upper limits for EDMs of diamagnetic systems (see Fig.~\ref{fig:EDM_all}). At present, the $^{199}$Hg atomic EDM measurement is the most sensitive, and its upper limit sets constraints to multiple sources of $\mathcal{CP}$ violation \cite{Graner2016}. Considering various potential contributions to an atomic EDM, an improved limit on other systems, like the $^{129}$Xe EDM $d_\mathrm{A} (^{129}\mathrm{Xe})$, will tighten these constraints. The theoretical results for $^{129}$Xe EDM are more accurate and reliable than those obtained for $^{199}$Hg EDM, therefore $^{129}$Xe has the potential to probe new physics \cite{Sakurai2019}.

Recently, new upper bounds on the $^{129}$Xe EDM using $^{3}$He comagnetometry and SQUID detection have been reported by a joint collaboration between the University of Michigan, the Technical University of Munich and the Physikalisch-Technische Bundesanstalt(PTB) \cite{Sachdeva2019} as well as another independent group with comparable sensitivities \cite{Allmendinger2019}, which are about five times smaller than the previous limit set in 2001 \cite{Rosenberry2001}. One of the challenges in both experiments is the comagnetometer frequency drift, which is several magnitudes larger than the expected frequency shift due to a potential $^{129}$Xe EDM  \cite{Gemmel2010}. One approach to correct for the impact of the comagnetometer drift on the measured $d_\mathrm{A} (^{129}\mathrm{Xe})$ is using a deterministic physical model to fit the comagnetometer frequency drift \cite{Allmendinger2019,Allmendinger2014}. However, the physical origin of the comagnetometer frequency instability is subject of a controversial debate \cite{Allmendinger2014a,Romalis2014}, which was inspired by another theoretical model and motivated the performance of recent experiments to substantiate the former criticism  \cite{Limes2019,Terrano2019}. Instead, in Ref.~\cite{Sachdeva2019} a phenomenological method was used, which does not need any physical model on the comagnetometer frequency drift, but a distinct pattern of electric fields with switching polarity. We will refer to that as the Pattern Combination (PC) method from here on. 

Here, we propose a new analysis based on a Global Phase Fitting (GPF) method, where the EDM value is estimated by a single fit to the comagnetometer phase development within one complete measurement. Besides an experimentally deduced EDM function as used in Ref.~\cite{Allmendinger2019}, allowing to analyse any electric field pattern, our GPF method uses a polynomial function to account for the comagnetometer frequency drift. Sec.~\ref{sec:principle} gives a short description of the basic principle of measuring the $^{129}$Xe EDM $d_\mathrm{A} (^{129}\mathrm{Xe})$ using comagnetometry. In addition, the PC method is introduced for comparison with the GPF method. The GPF method is elucidated in detail in Sec.~\ref{sec:GPF}, including the derivation of the Cramer-Rao Lower Bound (CRLB). The CRLB of the variance on the EDM value estimation using the GPF method is a factor of four smaller than that of the PC method. In Sec.~\ref{sec:results} we validate the GPF method with Monte-Carlo simulations and compare the results of the PC and GPF method using the experimental data obtained for Ref.~\cite{Sachdeva2019}. Eventually we recalculate the systematic uncertainties based on Ref.~\cite{Sachdeva2019} and derive a new upper limit for the permanent $^{129}$Xe EDM.

\section{$^{3}$He-$^{129}$Xe-COMAGNETOMETRY}
\label{sec:principle}
\subsection{Basic principle}
For $^{129}$Xe atoms stored in a cell permeated by a uniform magnetic field $\vec{B}$ and an electric field  $\vec{E}$, that is parallel to $\vec{B}$, their nuclear spin precesses at an angular frequency
\begin{equation} 
    \omega_{\mathrm{Xe}} = \left|\gamma_{\mathrm{Xe}} B + \frac{d_\mathrm{A} (^{129}\mathrm{Xe})E }{\hbar F_{\mathrm{Xe}} }\right|, 	\label{eqn:W_Xe}
\end{equation}
where $F_{\mathrm{Xe}}=1/2$ is the total angular momentum number and $\gamma_{\mathrm{Xe}}$ is the gyromagnetic ratio of $^{129}$Xe. The magnetic field $B$ in Eq.~(\ref{eqn:W_Xe}) becomes an interference term for directly calculating $d_\mathrm{A} (^{129}\mathrm{Xe})$ from $\omega_{\mathrm{Xe}}$. To overcome the experimental difficulties on controlling and measuring $\vec{B}$, comagnetometry was introduced with two collated species measured at the same time \cite{Rosenberry2001,Gemmel2010,Golub1994}. $^{3}$He is an ideal candidate for comagnetometry due to its potentially high SNR and a negligible EDM compared to $d_\mathrm{A} (^{129}\mathrm{Xe})$ \cite{Flambaum2012}. The weighted frequency difference between $^{129}$Xe atoms and $^{3}$He atoms is defined as 
\begin{equation}
    \omega_{\mathrm{co}}= \omega_{\mathrm{Xe}}-\frac{\gamma_{\mathrm{Xe}}}{\gamma_{\mathrm{He}}}\omega_{\mathrm{He}}, \label{eqn:Co_omega}
\end{equation}
and commonly named the comagnetometer frequency. Here $\omega_{\mathrm{He}} = |\gamma_{\mathrm{He}} B |$ is the spin precession frequency of $^{3}$He atoms with $\gamma_{\mathrm{He}}$ being its gyromagnetic ratio. Therefore, $\omega_{\mathrm{co}}$ can be written as
\begin{equation} 
    \omega_{\mathrm{co}}= \frac{2d_\mathrm{A} (^{129}\mathrm{Xe})}{\hbar}{\vec{E}} \cdot \hat{B},
    \label{eqn:C0_omega2}
\end{equation}
showing that $\omega_{\mathrm{co}}$  is independent of the magnitude of the background magnetic field but depends on its orientation relative to the applied electric field. The current measurement sensitivity of $\omega_{\mathrm{co}}$ is in the nHz range for a single measurement, while the comagnetometer frequency drift is at the \textmu Hz level, which causes a non-negligible systematic error \cite{Limes2019,Terrano2019,Thrasher2019}. Multiple physical models to describe the comagnetometer drift were proposed. The dominant terms thereby vary in different models. Furthermore, several parameters, such as the longitudinal relaxation time $T_1$ of the nuclear spins, used in these models are unknown or difficult to measure, making the frequency drift correction with a deterministic model inaccurate. By using a phenomenological model such as proposed here and in \cite{Sachdeva2019}, these currently unsolved difficulties can be omitted. 

\subsection{Parameters of two measurement campaigns}
\begin{figure*}[htpb]
    \centerline{\includegraphics[width=1\columnwidth]{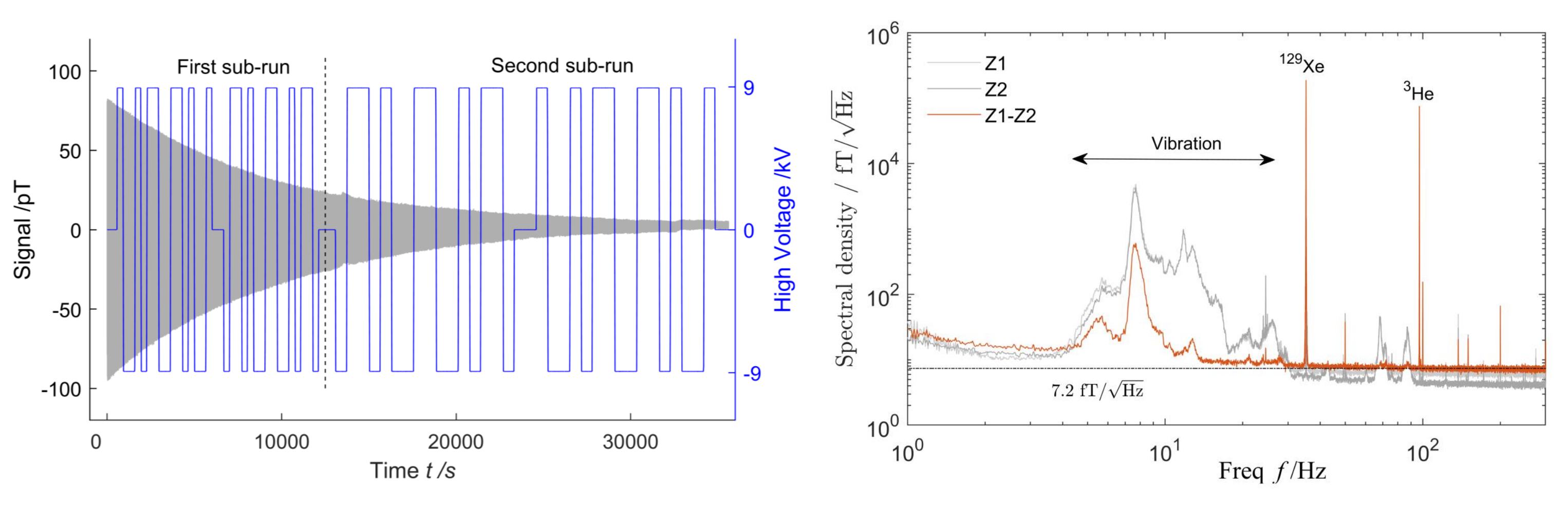}}
    \caption{ Left: The SQUID gradiometer $Z1-Z2$ signal (gray curve) and the modulated high voltage signal (blue line) of one run from the 2018 campaign. Right: The amplitude spectral density of data lasting 100~s from the starting of the first sub-run for two magnetometer channels ($Z1$ and $Z2$) and one software gradiometer ($Z1-Z2$). The white noise level of the  gradiometer is $\rho_\mathrm{\omega} \approx 7.2~\mathrm{fT}/\sqrt{\mathrm{Hz}}$. The variance of the white noise is $\sigma^2_\mathrm{\omega}=f_\mathrm{s}\rho_\mathrm{\omega}^2/2=(154~\mathrm{fT})^2$, with the sampling frequency $f_\mathrm{s}$ = 915.5245~Hz.}
    \label{fig:Noise}
\end{figure*}

The data used in our analysis were collected in the joint collaboration at the Berlin Magnetically Shielded Room (BMSR-2) facility at PTB Berlin. Table~\ref{tab:exp_para} summarizes the main experimental parameters of the two measurement campaigns carried out in 2017 and 2018, respectively. More details on the setup and process are given in Ref.~\cite{Sachdeva2019a}. The spin precession signal of the transverse magnetization of $^{3}$He and $^{129}$Xe was recorded by a dc-SQUID system with two channels ($Z1$,$Z2$). The high voltage and leakage current between the two electrodes of the cell were monitored. A background magnetic field $B_0$ in the range of 2.6 \textmu  T - 3 \textmu T was applied to shift $\omega_{\mathrm{Xe}}$ and $\omega_{\mathrm{He}}$ to 30 Hz - 36 Hz and 90 Hz - 98 Hz, respectively, which are well above the vibrational interference signals (see Fig.~\ref{fig:Noise}). In order to further decrease the impact of the vibrational noise, a software SQUID gradiometer ($Z1-Z2$) was used.  

\begin{table}[h]
    \centering
    \caption{Starting amplitude $A_0$, transverse relaxation time $T_2$, background magnetic field $B_0$ and segment length $t_\mathrm{s}$ of both measurement campaigns. }
    \begin{tabular}{c c c  }
    \hline\noalign{\smallskip}
         & \textbf{2017} & \textbf{2018} \\
    \noalign{\smallskip}\hline\noalign{\smallskip}
    $A_\mathrm{0,Xe}$~/pT & 30-40  & 70-80 \\
    $A_\mathrm{0,He}$~/pT & 4-5  & 20-25 \\
    $T_2^\mathrm{Xe}$~/s & 6000-7000 & 8000-9000 \\
    $T_2^\mathrm{He}$~/s & 7000-8000 & 8000-9000  \\
    $B_0$~/\textmu T  & 2.6 & 3  \\
    $t_\mathrm{s}$~/s  & 400 or 800 & 100-800 \\
    \noalign{\smallskip}\hline
    \end{tabular}
    \label{tab:exp_para}
\end{table}

The left panel of Fig.~\ref{fig:Noise} shows the raw SQUID gradiometer signal in pT (gray) of one run from the 2018 campaign lasting 35000~s exemplarily. This run comprises two so called sub-runs with 36 segments each. A segment is defined as the time of constant electric field. For the two sub-runs shown in Fig.~\ref{fig:Noise}, the segments last 300~s and 600~s, respectively. The first sub-run ranging from 50~s to 12400~s is used as an example in the data analysis section.   
\subsection{PC method}
As mentioned above, one approach to mitigate the effect of the comagnetometer frequency drift is repetitively reversing the direction of the electric field $\vec{E}$. This allows to separate the impact of $d_\mathrm{A} (^{129}\mathrm{Xe})$ on $\omega_{\mathrm{co}}$ from other interference terms. The $E$ modulation method has been applied in diverse EDM experiments with varied modulation patterns \cite{Bishof2016,Baker2006}. For the PC method, the common $E$ pattern for one sub-run consists of 36 segments with an equal time interval $t_\mathrm{s}$, and the sign of $\vec{E}$ changes according to the following sequence $\pm$[0 + - - + - + + - - + + - + - - + 0, 0 - + + - + - - + + - - + - + + - 0]. The segments of zero voltage were added to allow for systematic error studies.

The PC method determines the EDM value from averaging the comagnetometer frequencies $\omega_{\mathrm{co}}$ from $2^n$ ($n \in \mathbb{N}$) consecutive segments omitting those with zero voltages. This pattern is constructed to cancel the effect of the comagnetometer frequency drift up to $n-1$ order when parametrized in polynomials. The effect of the higher order (above $n-1$) drift dependency imposing a false EDM on each sub-run is deduced by applying polynomial fits to all $\omega_\text{co}$ within the sub-runs, leading to a correction for the EDM and an additional systematic uncertainty (for more details see Ref.~\cite{Sachdeva2019a}).

\section{GLOBAL PHASE FITTING METHOD}
\label{sec:GPF}
\begin{figure}[htpb]
    \centerline{\includegraphics[width=0.6\columnwidth]{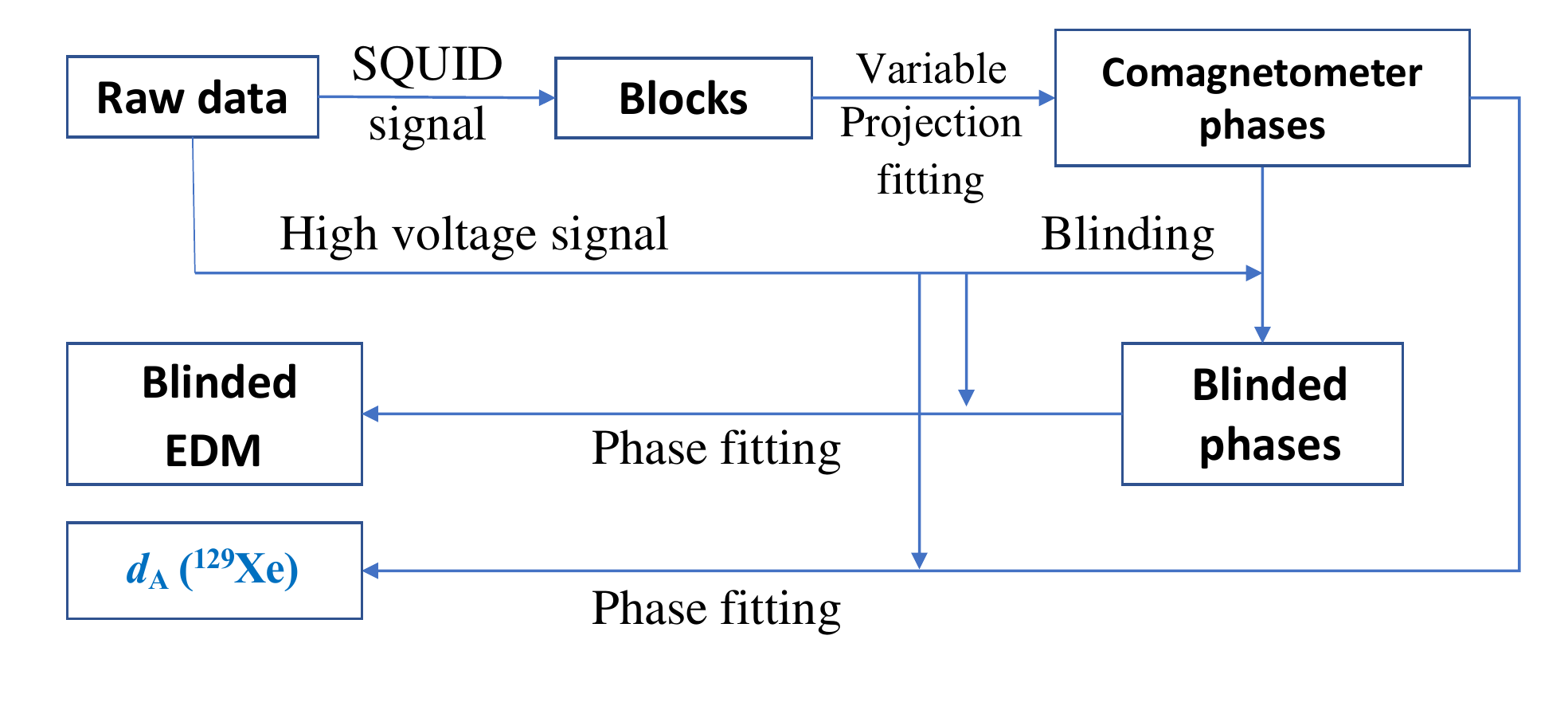}}
    \caption{ \label{fig:Schematic} The schematic process of the GPF method. }
\end{figure}

The general data-processing procedure for the GPF method is illustrated in Fig.~\ref{fig:Schematic}. For this method, the raw SQUID data of a sub-run is cut into continuous blocks of equal length. Each block data is fitted to deduce precession phases of both species $^{3}$He and $^{129}$Xe (see Sec.~\ref{subsec:phase}) and the continuous comagnetometer phase is derived for each block (see Sec.~\ref{subsec:continuous_phase}). For data blinding an additional phase, bound to the measured high voltage signal, can be added to the comagnetometer phase at this point (see Sec.~\ref{subsec:fit_EDM}). The EDM value is acquired by fitting the blinded comagnetometer phases using a polynomial function together with a constructed function comprising the phase evolution introduced by a hypothetical $^{129}$Xe EDM. The unblinded EDM result is obtained by reanalyzing the raw comagnetometer phases, as illustrated in Fig.~\ref{fig:Schematic}.

\subsection{The phase of each block}
\label{subsec:phase}
The block length $t_{\mathrm{b}}$ is a free parameter with a suitable range from 1~s to 20~s, being short enough to exclude the amplitude decay and frequency drift, and long enough to perform the fit for our data  \cite{Sachdeva2019}. The SQUID data in each block are fitted to the function
\begin{equation}
    \label{eqn:sincos_fit}
    y= a_{\mathrm{Xe}}\sin(\omega_{\mathrm{Xe}}t)+b_{\mathrm{Xe}}\cos(\omega_{\mathrm{Xe}}t)     +a_{\mathrm{He}}\sin(\omega_{\mathrm{He}}t)+b_{\mathrm{He}}\cos(\omega_{\mathrm{He}}t)+a_{i}\sin(\omega_{i}t)+b_{i}\cos(\omega_{i}t)+c+d \cdot t
    ,
\end{equation} 
where $a_{\mathrm{Xe/He}/i},b_{\mathrm{Xe/He}/i},\omega_{\mathrm{Xe/He}},c$,~and~$d$ are the fit parameters and $\omega_{i=1,2,3,4}=2\pi \times 50i$~s$^{-1}$  represent the power frequency and its harmonics. The constant and linear terms $c$ and $d\cdot t$ describe the background magnetic field and its small drift as seen by the SQUID. The variable projection (VP) method is applied \cite{Golub2003}, where the nonlinear parameters $\omega_{\mathrm{Xe/He}}$ are estimated separately from the linear parameters $a_{\mathrm{Xe/He}/i},b_{\mathrm{Xe/He}/i},c,$~and~$d$. To minimize the correlation between the fit terms in Eq.~(\ref{eqn:sincos_fit}), the time of each block is assigned to be symmetrical around zero from $-t_{\mathrm{b}}/2$ to $t_{\mathrm{b}}/2$. 

\begin{figure}[htpb]
    \centerline{\includegraphics[width=0.6\columnwidth]{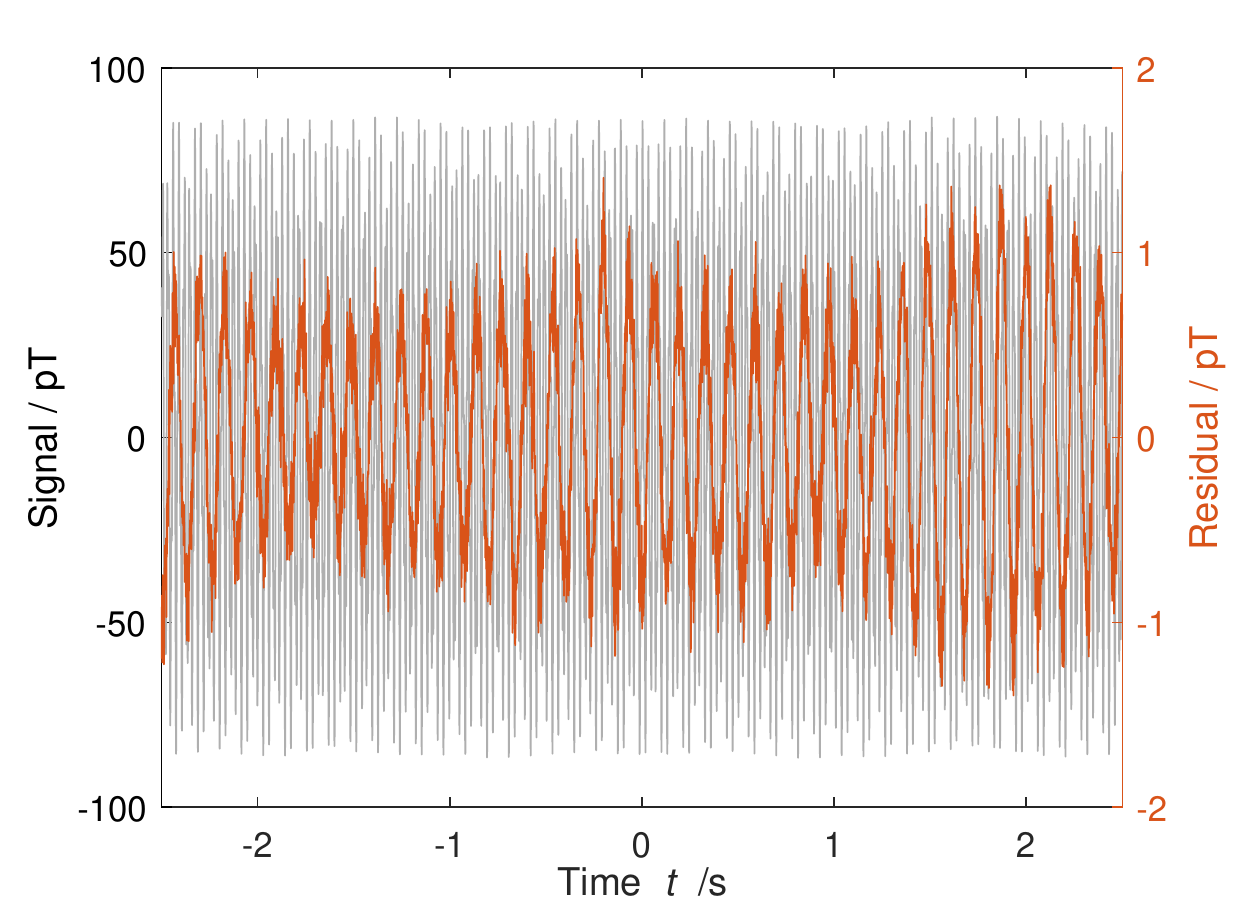}}
    \caption{The gradiometer signal (gray) for one block with $t_\mathrm{b}=5 $~s and the residual of the fit (red). }
    \label{fig:block_fit}
\end{figure}

Fig.~\ref{fig:block_fit} shows the raw SQUID data of a 5~s block from the start of the exemplary sub-run and the residual of the fit to this data. The residual is dominated by the mechanical vibration in the frequency range of 4~Hz - 25~Hz as shown in the right plot of Fig.~\ref{fig:Noise}. We can assume approximate orthogonality between the precession signal and the vibrational noise of our setup. Therefore, the error on the fit parameter values caused by the latter one is negligible compared to that caused by the white noise, although its integrated power is much larger than the white noise power. This was validated with Monte-Carlo simulations using the recorded vibrational noise (see Appendix~\ref{subsec:Vibra}). The phase of each species for the block~$k$ in the range of $[-\pi, \pi)$ can be obtained by  
\begin{equation} 
    \label{eqn:phase}
    \phi_{\mathrm{m}}^k=\mathrm{Arg}(a_{\mathrm{m}}^k+i \cdot b_{\mathrm{m}}^k),
\end{equation}
where Arg is the function to get the principle argument of a complex number, $i$ is the imaginary unit and m = Xe or He. Note that due to the time centering, the estimated phase $\phi^k$ is referred to the middle time of each block $t_k=(k-1/2)t_\mathrm{b}$. The time interval of the block $k$ is defined as $(t_{k-1},t_k)$. The parameter uncertainties $\delta a_{\mathrm{m}}^k$ and $\delta b_{\mathrm{m}}^k$ are estimated from the covariance matrix of the fit
\begin{equation} 
   \label{eqn:uncer}
   \mathrm{Cov}=\frac{\vec{r}^\prime \cdot \vec{r}} {\nu (\tens{J}^\prime \cdot  \tens{J})^{-1}},
\end{equation}
where $\vec{r}$ is the residual, $\nu$ is the degrees of freedom and $\tens{J}$ is the Jacobian matrix. The standard deviation of the derived phase $\delta \phi^k_\mathrm{Xe/He}$ is
\begin{equation} 
    \label{eqn:phase_std}
    \delta\phi_{\mathrm{m}}^k=\frac{\sqrt{(a_{\mathrm{m}}^k \cdot \delta b_{\mathrm{m}}^k)^2+(b_{\mathrm{m}}^k \cdot \delta a_{\mathrm{m}}^k)^2 - 2a_{\mathrm{m}}^k b_{\mathrm{m}}^k \mathrm{Cov}(a_{\mathrm{m}}^k,b_{\mathrm{m}}^k)}}{(a_{\mathrm{m}}^k)^2+(b_{\mathrm{m}}^k)^2}.
\end{equation}
Eq.~(\ref{eqn:uncer}) assumes that the residual $\vec{r}$ stems from the wideband white noise, which is a conservative approach for our case since the main signal in the residuals is the narrowband vibrational noise, leading to an overestimation of the uncertainty $\delta\phi^k_{\mathrm{m}}$. However, the ratio between $\delta\phi^k_\mathrm{m}$ for different blocks reflects the decaying SNR. Therefore, these estimated uncertainties are used as weights in the subsequent GPF routine.  
\subsection{The accumulated comagnetometer phase}
\label{subsec:continuous_phase}
The accumulated phase $\Phi^k_{\mathrm{m}}$ in a block $k$ of the continuously precessing spins is the sum of the wrapped phase $\phi^k_{\mathrm{m}}$  and a multiple of $2\pi$ 
\begin{equation} 
    \label{eqn:Phase}
    \Phi_{\mathrm{m}}^{k}=\phi_{\mathrm{m}}^{k}+2 \pi n_{\mathrm{m}}^{k},
\end{equation}
where the integer $n_{\mathrm{m}}^{k}$  is determined as
\begin{equation} 
    \label{eqn:integer}
    n_{\mathrm{m}}^{k}=\frac{\Phi_{\mathrm{m}}^{k-1}+\omega_{\mathrm{m}}^{k-1} t_{\mathrm{b}}}{2 \pi},
\end{equation}
rounded to the lower integer and $n_{\mathrm{m}}^{1}=0$. Here, the frequencies $\omega_{\mathrm{m}}^{k-1}$  are obtained by the fit of the block $k$ using Eq.~(\ref{eqn:sincos_fit}). If $\Phi_{\mathrm{m}}^{k}-(\Phi_{\mathrm{m}}^{k-1}+\omega_{\mathrm{m}}^{k-1} t_{\mathrm{b}})$ is either $>\pi$ or $<-\pi$, $n_{\mathrm{m}}^{k}$ is incremented or decremented by one, respectively, to ensure a continuous phase evaluation. The standard deviation of the accumulated phase  $\delta\Phi_{\mathrm{m}}^{k}$  is equal to $\delta\phi_{\mathrm{m}}^{k}$ as Eq.~(\ref{eqn:Phase}) does not introduce any additional uncertainty. According to Eq.~(\ref{eqn:Co_omega}), the evolved comagnetometer phase $\Phi_{\mathrm{co}}^{k}$ for each block $k$ is determined by
\begin{equation} 
    \label{eqn:co_phase}
    \Phi_{\mathrm{co}}^{k}=\Phi_{\mathrm{Xe}}^{k}-\frac{\gamma_{\mathrm{Xe}}}{\gamma_\mathrm{He}}\Phi_{\mathrm{He}}^{k}.
\end{equation}

\subsection{The fitted EDM value}	
\label{subsec:fit_EDM}
By integrating Eq.~(\ref{eqn:C0_omega2}), the accumulated phase due to a hypothetical $^{129}$Xe EDM $d_\mathrm{h}$ at the block $k$ is 
\begin{equation}
    \label{eqn:Phi_EDM}
    \Phi_{\mathrm{EDM}}^k= t_b \frac{2d_\mathrm{h}}{\hbar}\sum_{i=1}^{k}({\vec{E}_i} \cdot \hat{B}),
\end{equation}
where $\vec{E}_i$ is the average electric field within the block $i$. By replacing  $d_\mathrm{h}$ with a computer-generated pseudo-random EDM value $d_\mathrm{bias}$, the bias phase $ \Phi_{\mathrm{bias}}^k$ is calculated and then used to blind the comagnetometer phase $\Phi_{\mathrm{co,b}}^k = \Phi_{\mathrm{co}}^k+\Phi_{\mathrm{bias}}^k$ in order to avoid operator induced bias during process optimization. The value of $d_{\mathrm{bias}}$   has been saved in an independent file in a binary format and  $\Phi_{\mathrm{co,b}}^k$ was used for later data analysis.

\begin{figure}[htpb]
    \centerline{\includegraphics[width=0.7\columnwidth]{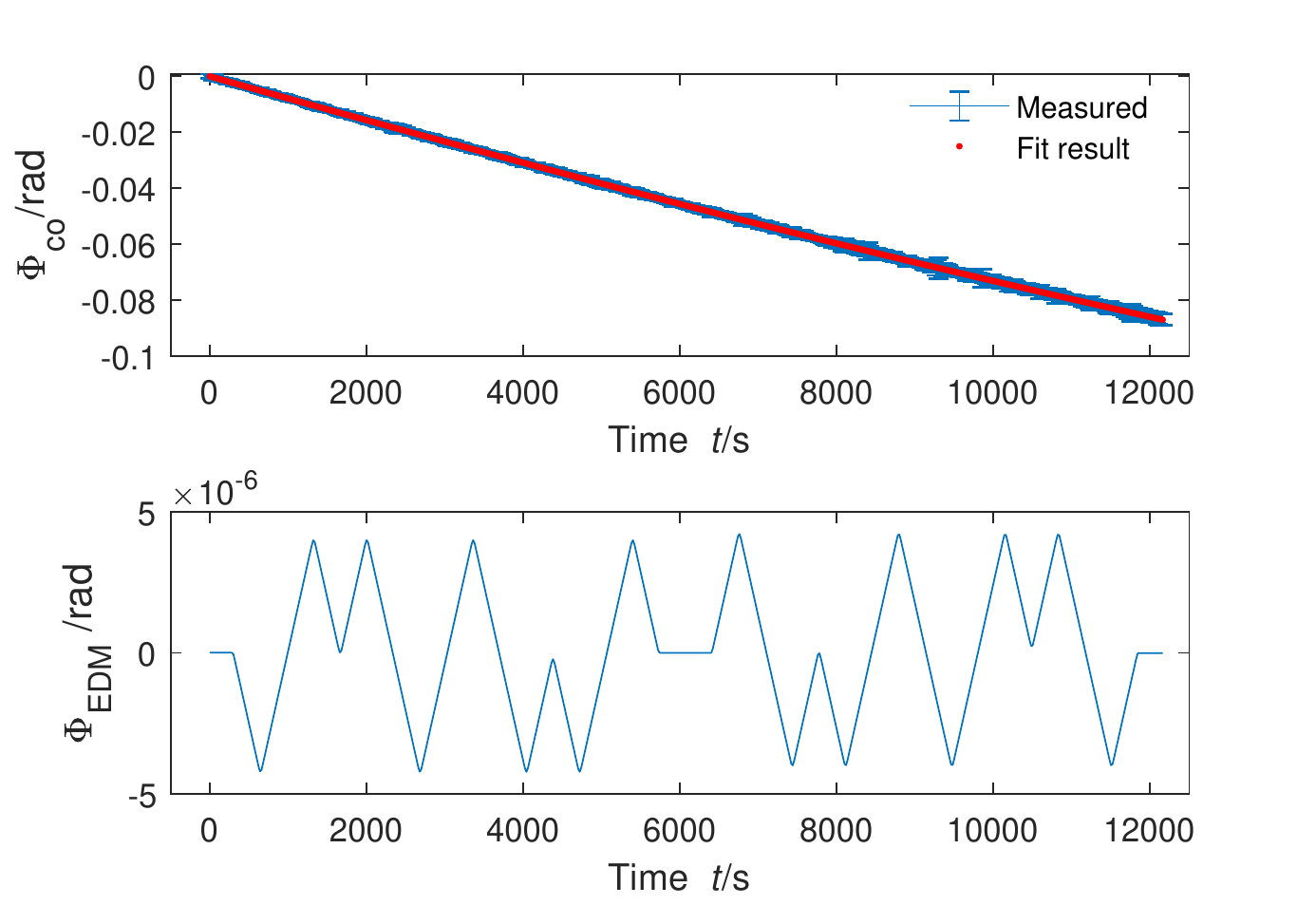}}
    \caption{ \label{fig:Phi} The comagnetometer phase $\Phi_{\mathrm{co}}^k$  and the data of the fit $\Phi_{\mathrm{fit}}^k$  (top) as well as the EDM function  $\Phi_{\mathrm{EDM}}^k$  constructed from the measured electric field (bottom) with $d_\mathrm{h} = 1 \times 10^{-27}~e~\mathrm{cm}$ for the exemplary sub-run. }
\end{figure}

The measured phase $\Phi_{\mathrm{co}}^k$ originates not only from the potential $^{129}$Xe EDM, but also from other sources such as chemical shift \cite{Limes2019,Sachdeva2019a}. These contributions are phenomenologically parametrized by a polynomial of $g$th order \cite{Tullney2013}. Hence, the comagnetometer phase is fitted with the function 
\begin{equation}
    \label{eqn:GPF}
    \Phi_{\mathrm{fit}}^k= a \Phi_{\mathrm{EDM}}^k + p_0 + p_1 \tilde{P}_1(t_k) + p_2\tilde{P}_2(t_k) + \cdots + p_g \tilde{P}_g(t_k) ,
\end{equation}
where $a, p_0, p_1, p_2, …, p_g$ are the global fit parameters. Here the time series $t_k$ are normalized to the interval [0,1] and shifted Legendre polynomials $\tilde{P}_n(t_k)$ are applied to decrease the correlation between polynomial coefficients \cite{Refaat2009}. The fit was conducted by using the iterative least squares estimation method with the built-in function $nlinfit$ in MATLAB.  Thereby the inverse values of the phase variances $(\delta\Phi_{\mathrm{co}}^k)^2$ are used as weights. Fig.~\ref{fig:Phi} shows the comagnetometer phase $\Phi_{\mathrm{co}}^k$, the fit phase  $\Phi_{\mathrm{fit}}^k$, and the EDM function $\Phi_{\mathrm{EDM}}^k$ constructed from the measured $E$-field pattern of the exemplary sub-run. 
To determine the order needed for the polynomial function in Eq.~(\ref{eqn:GPF}), we apply an $F$-test where the significance of adding $q$ terms to the fitting function with $g$ terms was evaluated by the integral probability
\begin{equation}
    \label{eqn:P}
        P_{g,g+q}=\int_{0}^{F_{g,g+q}}P_\mathrm{F}(F;q,N-g-q)dF,
\end{equation}
where $P_\mathrm{F}$ is the probability density function of the $F$-distribution and $N$ is the number of data points \cite{Bevington1992}. The upper bound of the integral is
\begin{equation}
    \label{eqn:F}
        F_{g,g+q}=\frac{(N-g-q)(\chi_g^2-\chi_{g+q}^2)}{q \cdot \chi_{g+q}^2}.
\end{equation}
The order of the fit was defined sufficient when  $P_{g,g+1}$  as well as $P_{g,g+2}$ are both smaller than a chosen threshold of  $P_{\mathrm{min}}$. 

The atomic EDM of $^{129}$Xe is calculated from the fit parameter $a$ as
\begin{equation}
    \label{eqn:d_Xe}
        d_\mathrm{A} (^{129}\mathrm{Xe})=a \cdot d_\mathrm{h}.
\end{equation}
The correlated uncertainties of the parameters are determined as the square root of the reciprocal of the diagonal of the covariance matrix, which inherently includes the uncertainty of the correlations between $a$ and polynomial parameters. The influences of these correlations to the estimation of $a$ are small due to the orthogonality between the constructed function  $\Phi_{\mathrm{EDM}}^k$  and the polynomial function of the order up to $n-2$ where $2^n$ is the number of nonzero high voltage segments. The correlation matrix for the exemplary sub-run (see Fig.~\ref{fig:Noise}) is given in Table~\ref{tab:corr_GPF_fit}. In this case, the correlations between the EDM parameter $a$ and the polynomial coefficients are significantly smaller than 1, but nonzero, since the polynomials of higher than 3rd order are not orthogonal to  $\Phi_{\mathrm{EDM}}^k$. The derived uncertainty is in good agreement with the result using the log profile likelihood method. We also applied the linear regression method with the model in Eq.~(\ref{eqn:GPF}) and obtained consistent results.  

\begin{table}[h]
    \centering
    \caption{Correlation matrix of the first sub-run for the fit with a 7th order polynomial and the block length $t_b$=5~s.}
    \begin{tabular}{c c c c c c c c c c}
    \hline\noalign{\smallskip}
          & $a$ & $p_0$ & $p_1$  & $p_2$  & $p_3$ & $p_4$ & $p_5$ & $p_6$  & $p_7$ \\
    \noalign{\smallskip}\hline\noalign{\smallskip}
    $a$   & \textbf{1.0} & \textbf{0.0} & \textbf{0.0} & \textbf{0.0} & \textbf{0.0} & \textbf{0.2} & \textbf{0.0} & \textbf{0.2} & \textbf{0.2} \\
    $p_0$ & \textbf{0.0} & 1.0 & 0.5 & 0.2 & 0.1 & 0.0 & 0.1 & 0.1 & 0.0\\
    $p_1$ & \textbf{0.0} & 0.5 & 1.0 & 0.5 & 0.2 & 0.1 & 0.1 & 0.1 & 0.0 \\
    $p_2$ & \textbf{0.0} & 0.2 & 0.5 & 1.0 & 0.5 & 0.2 & 0.1 & 0.0 & 0.0 \\
    $p_3$ & \textbf{0.0} & 0.1 & 0.2 & 0.5 & 1.0 & 0.5 & 0.2  & 0.1 & 0.0 \\
    $p_4$ & \textbf{0.2} & 0.0 & 0.1 & 0.2 & 0.5 & 1.0 & 0.5  & 0.2 & 0.1 \\
    $p_5$ & \textbf{0.0} & 0.1 & 0.1 & 0.1 & 0.2 & 0.5 & 1.0 & 0.5 & 0.1 \\
    $p_6$ & \textbf{0.2} & 0.1 & 0.1 & 0.0 & 0.1 & 0.2  & 0.5 & 1.0 & 0.5 \\
    $p_7$ & \textbf{0.2} & 0.0 & 0.0 & 0.0 & 0.0 & 0.1 & 0.1 & 0.5 & 1.0 \\
    \noalign{\smallskip}\hline
    \end{tabular}
    \label{tab:corr_GPF_fit}
\end{table}

\subsection{The modified Allan deviation}	
\begin{figure}[htpb]
    \centerline{\includegraphics[width=0.6\columnwidth]{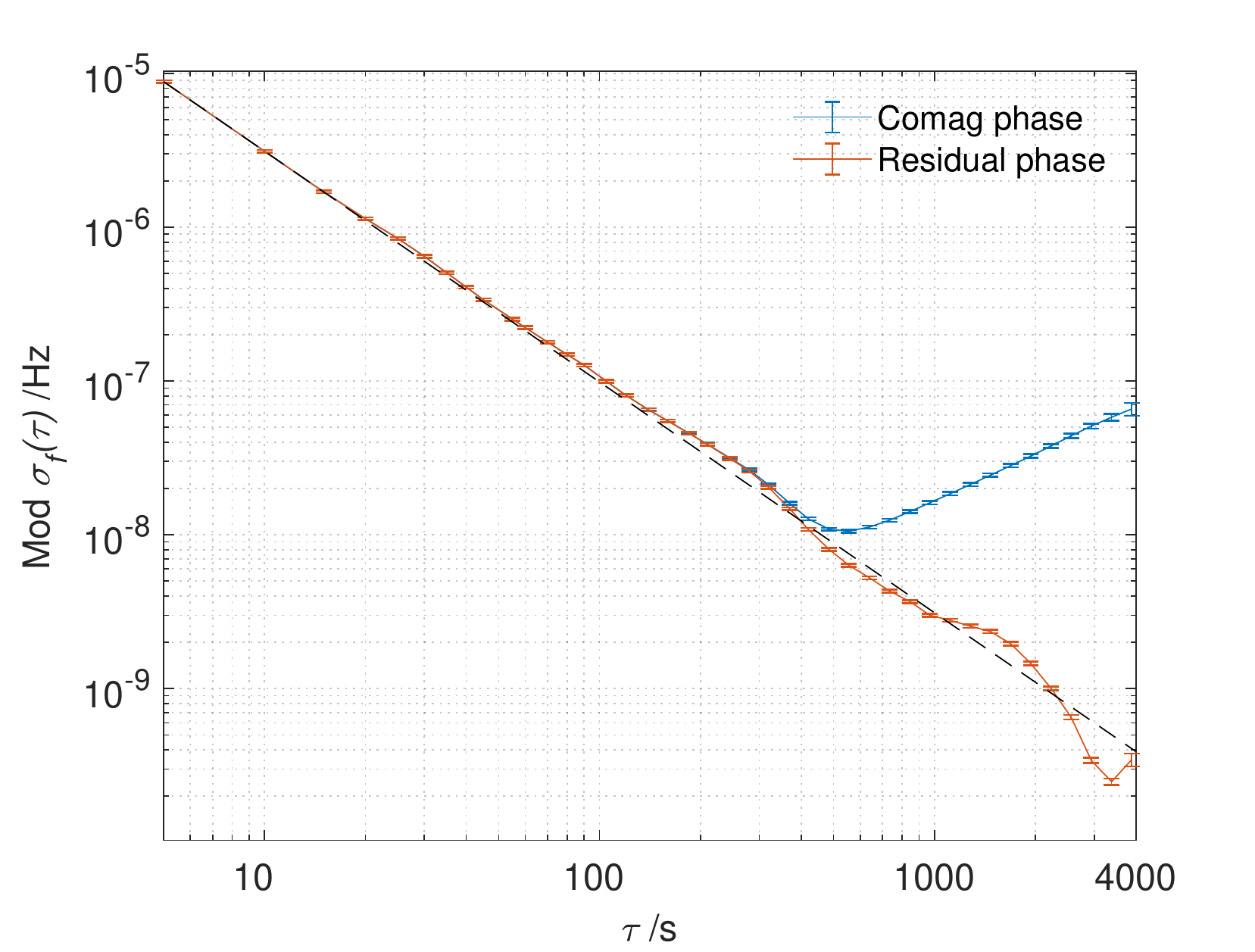}}
    \caption{ \label{fig:Allan} The modified Allan deviation and its error bar of the accumulated comagnetometer phase and the residual phases for the fit with a 7th order polynomial. To fulfill the MAD statistics criteria \cite{Allan1981}, only data are shown for integration time $\tau < 4000$~s.}
\end{figure}

The modified Allan deviation (MAD) is an established tool to evaluate the low-frequency drift of a time series of phases $\Phi$, which is defined as  
\begin{equation}
    \label{eqn:Allan}
        \sigma_f(\tau)=\frac{1}{2\pi}\sqrt{\frac{\sum\limits^{P-3n-1}_{j=1}\left(\sum\limits_{k=j}^{j+n-1}\Phi^{k+2n}-2\Phi^{k+n}+\Phi^{n}\right)^2}{2n^2\tau^2(P-3n+1)}},
\end{equation}
where the integration time $\tau$ is $n$ times the block length $t_\mathrm{b}$, and the total measurement time $T$ is subdivided into $P$ time intervals of equal length $\tau$, such that $P\tau \approx T $ \cite{Allan1981}. As an example, the MAD of the exemplary sub-run is plotted in Fig.~\ref{fig:Allan}. $\sigma_f$ of $\Phi_{\mathrm{co}}^k$  reaches the minimum at the integration time $\tau$ of 550~s and then increases due to the comagnetometer frequency drift. For the residual phase $\Phi_{\mathrm{co}}^k-\Phi_{\mathrm{fit}}^k$ of this exemplary sub-run, the MAD decreases with increasing integration time according to $\sigma_f \propto \tau^{-3/2}$ (dashed line in Fig.~\ref{fig:Allan}) over the considered range, down to 0.4~nHz. This behavior is an indicator that the comagnetometer phase $\Phi_{\mathrm{co}}^k$  is adequately described by the fit model of Eq.~(\ref{eqn:GPF}), since the residual is dominated by white phase noise.

\subsection{The theoretical statistical uncertainty bound}	
The theoretical limit of the $^{129}$Xe EDM uncertainty can be derived as the CRLB, which also provides insights into optimizing experimental parameters. For the sake of simplicity, only a single species spin-precession signal is considered and its amplitude is assumed to be a constant over the whole sub-run. For the GPF method, $d_\mathrm{A} (^{129}\mathrm{Xe})$ is estimated with two steps: The VP fitting to obtain the phase of each block and global phase fitting of the sub-runs. Therefore, the overall CRLB is the combination of the results of these two fits. 

For the phase $\phi$ of a sinusoid embedded in white Gaussian noise (WGN) observed over one block with time being symmetrically around 0~s, the CRLB is 
\begin{equation}
    \label{eqn:CRLB_phase}
        \mathrm{var}(\hat{\phi}) \geq \frac{2\sigma_{\mathrm{w}}^2}{A^2 N},
\end{equation}
where $\sigma_{\mathrm{w}}^2$  is the variance of the WGN, $A$ the amplitude and $N$ the number of data points in one block \cite{KAY1993}. The CRLB for the parameters in the fit model Eq.~(\ref{eqn:GPF}) is the reciprocal of the Fisher information matrix 
\begin{equation}
    \label{eqn:Fisher_GPF}
        \tens{I} =      
        \begin{pmatrix}
              \sum\limits_{k=1}^{JM}\frac{ (\Phi_{\mathrm{EDM}}^k)^2}{\delta \phi_k^2} & \cdots & \cdots & \sum\limits_{k=1}^{JM}\frac{ \Phi_{\mathrm{EDM}}^k t_k^g}{\delta \phi_k^2} \cr
            \sum\limits_{k=1}^{JM}\frac{\Phi_{\mathrm{EDM}}^k}{\delta \phi_k^2} & \sum\limits_{k=1}^{JM}\frac{1}{\delta \phi_k^2} & \cdots & \vdots \cr
            \vdots & \vdots & \ddots & \vdots \cr
            \sum\limits_{k=1}^{JM}\frac{\Phi_{\mathrm{EDM}}^k t_k^g}{\delta \phi_k^2} & \sum\limits_{k=1}^{JM}\frac{t_k^g}{\delta \phi_k^2} & \cdots &\sum\limits_{k=1}^{JM}\frac{t_k^g t_k^g}{\delta \phi_k^2} \cr
       \end{pmatrix}  ,
\end{equation}
where $M$ is the number of segments in one sub-run, and $J$ is the number of blocks in one segment. For the sake of simplicity, the standard polynomial is used in the fit model Eq.~(\ref{eqn:GPF}). Assuming $\sum\limits_{k=1}^{JM} \Phi_{\mathrm{EDM}}^k t_k^i = 0 $ for $i$ going from 0 to $g$ and the phase uncertainty $\delta\phi$  is a constant, the considered CRLB can be simplified to the so called ideal or uncorrelated CRLB as
\begin{equation}
    \label{eqn:CRLB_EDM}
    \mathrm{var}(\hat{d}_\mathrm{A} (^{129}\mathrm{Xe}))_\mathrm{GPF} \geq  \frac{\delta \phi^2}{\sum\limits_{k=1}^{JM}(\Phi_{\mathrm{EDM}}^k)^2}.
\end{equation}

By substituting Eqs.~(\ref{eqn:Phi_EDM}), (\ref{eqn:CRLB_phase}) and (\ref{eqn:Fisher_GPF}) into Eq.~(\ref{eqn:CRLB_EDM}), and exploiting the periodic property of the constructed EDM function (see Fig.~\ref{fig:Phi}), for our case, $\sum\limits_{k=1}^{JM} (\Phi_{\mathrm{EDM}}^k)^2 = M \sum\limits_{k=1}^{J} (\Phi_{\mathrm{EDM}}^k)^2 $, the overall CRLB for $d_\mathrm{A} (^{129}\mathrm{Xe})$  becomes 
\begin{align}
    \label{eqn:CRLB_EDM2}
    \mathrm{var}(\hat{d}_\mathrm{A} (^{129}\mathrm{Xe}))_\mathrm{GPF} & \geq  \left. \frac{2\sigma_{\mathrm{w}}^2}{A^2 N} \right/ \left( \left(\frac{2|E|t_b}{\hbar} \right)^2 M \sum\limits_{k=1}^{J} k^2 \right)  \\
    & \geq \frac{\sigma_{\mathrm{w}}^2}{A^2}\left(\frac{\hbar}{2|E|}\right)^2\frac{6M^2 \delta t}{T^3}, 
\end{align}
where $T=MJN\Delta t$ is the total measurement time and $\Delta t =1/f_\mathrm{s}$ is the sampling interval. Note that the number of segments $M$ should be large enough to ensure the orthogonality between $\Phi_{\mathrm{EDM}}^k$ and the polynomial functions. In case of an exponentially decaying amplitude $A$ of the precession signal, the CRLB has to be calculated with Eq.~(\ref{eqn:Fisher_GPF}).
For the PC method, the CRLB on the $^{129}$Xe EDM for $ M$ segments is derived in Ref.~\cite{Sachdeva2019a} as 
\begin{equation}
    \label{eqn:CRLB_PC}
   \mathrm{var}(\hat{d}_\mathrm{A} (^{129}\mathrm{Xe}))_\mathrm{PC}\geq  \frac{\sigma_{\mathrm{w}}^2}{A^2}\left(\frac{\hbar}{2|E|}\right)^2\frac{24M^2 \delta t}{T^3}.
\end{equation}

The PC method applies linear fits to the comagnetometer phases within one segment to derive the comagnetometer frequency of each segment, which requires the addition of an interception term as a starting phase, increasing the variance by a factor of four compared to a linear fit without interception term. In the GPF method the accumulated comagnetometer phases within one sub-run are analyzed in a single fit, therefore the uncertainty does not increase as the interception term is orthogonal to the EDM function (see Eq.~(\ref{eqn:Fisher_GPF})). Furthermore, the PC method requires the unweighted average of at least four segment frequencies, which increases its statistical uncertainty even further. 

\section{Results}
\label{sec:results}
A Monte-Carlo study was conducted to confirm that the GPF method can reach the higher sensitivity as shown by the CRLB compared to the PC method. Later, the GPF method was used to obtain the $^{129}$Xe EDM from the data set as taken in Ref.~\cite{Sachdeva2019} using the same channel and block length for analysis. As there were data sets in the 2017 and 2018 campaigns which were not useable with the PC method but could be analyzed with the GPF method, we gathered all data and optimized the analysis parameters to obtain the minimum uncertainty from the data. Ultimately, an improved upper limit of the $^{129}$Xe EDM was derived using the unblinded data.    
\subsection{Monte-Carlo tests}
The accumulated phase of each spin species for the sampling point $j$ was generated as 
\begin{equation}
    \label{eqn:Phi_Xesyn}
     \Phi_{\mathrm{Xe,syn}}^j =\int_{0}^{t_j} \gamma_\mathrm{Xe}B(t)+ 2 \pi (f_{\mathrm{lin}}^{\mathrm{Xe}}+u^\mathrm{Xe}e^{-t/T_1^\mathrm{Xe}}+f_\mathrm{EDM}) dt,
\end{equation}
\begin{equation}
    \label{eqn:Phi_Hesyn}
   \Phi_{\mathrm{He,syn}}^j = \int_{0}^{t_j} \gamma_\mathrm{He}B(t)+ 2 \pi (f_{\mathrm{lin}}^{\mathrm{He}}+u^\mathrm{He}e^{-t/T_1^\mathrm{He}}) dt ,  
\end{equation}
where the drift of the background field $B(t)$ was parametrized with a 4th order polynomial. $f_{\mathrm{lin}}^\mathrm{Xe/He} $ represent the frequency shifts caused by the chemical shift and Earth’s rotation. $u^\mathrm{Xe/He}$ are the drift amplitudes of the respective precession frequencies. The frequency drift was modeled as exponentially decaying functions with the characteristic time of $T_1$ \cite{Limes2019,Terrano2019,Thrasher2019}. Thereby it was assumed that $T_1$ is larger than $T_2$ and its range is listed in Table~\ref{tab:Syn_para}. $f_\mathrm{EDM}$ is the frequency shift due to the coupling of a synthetic EDM $d_\mathrm{syn}$ with the electric field according to Eq.~(\ref{eqn:C0_omega2}). Substituting Eqs.~(\ref{eqn:Phi_Xesyn}) and (\ref{eqn:Phi_Hesyn}) into Eq.~(\ref{eqn:co_phase}) results in the synthetic comagnetometer phase, whose time dependence is designed to mimic the measured data (for details see 
Appendix~\ref{subsec:phase_drift}). The exponentially decaying spin precession signals of $^{129}$Xe and $^{3}$He atoms can be described by
\begin{equation}
    \label{eqn:SQUID_syn}
    V_{\mathrm{Xe/He}}^j = A_0^{\mathrm{Xe/He}} e^{-t_j/T_2^{\mathrm{Xe/He}}}\sin{\Phi_{\mathrm{Xe/He,syn}}^j}
\end{equation}
with $t_j=j\Delta t$ , which is the time for the sampling point $j$. 

\begin{table}[h]
    \centering
    \caption{ The range of the parameter values used for generating synthetic spin precession data for Monte-Carlo simulations.}
    \begin{tabular}{c c | c c}
    \hline\noalign{\smallskip}
     Para. & Range & Para.  & Range \\
    \noalign{\smallskip}\hline\noalign{\smallskip}
    $u^{\mathrm{He}}$    &  3.5-4.5 \textmu Hz   &  $T_1^{\mathrm{He}}$  & 9000-14000 s   \\
    $u^{\mathrm{Xe}}$    &   9-11 \textmu Hz  &  $T_1^{\mathrm{Xe}}$  & 9000-14000 s   \\
    $f_{\mathrm{lin}}^{\mathrm{He}}$  &  4-10 \textmu Hz      & $f_{\mathrm{lin}}^{\mathrm{Xe}}$ & 4-10 \textmu Hz     \\
    \noalign{\smallskip}\hline
    \end{tabular}
    \label{tab:Syn_para}
\end{table}

\begin{figure}[htpb]
    \centerline{\includegraphics[width=0.6\columnwidth]{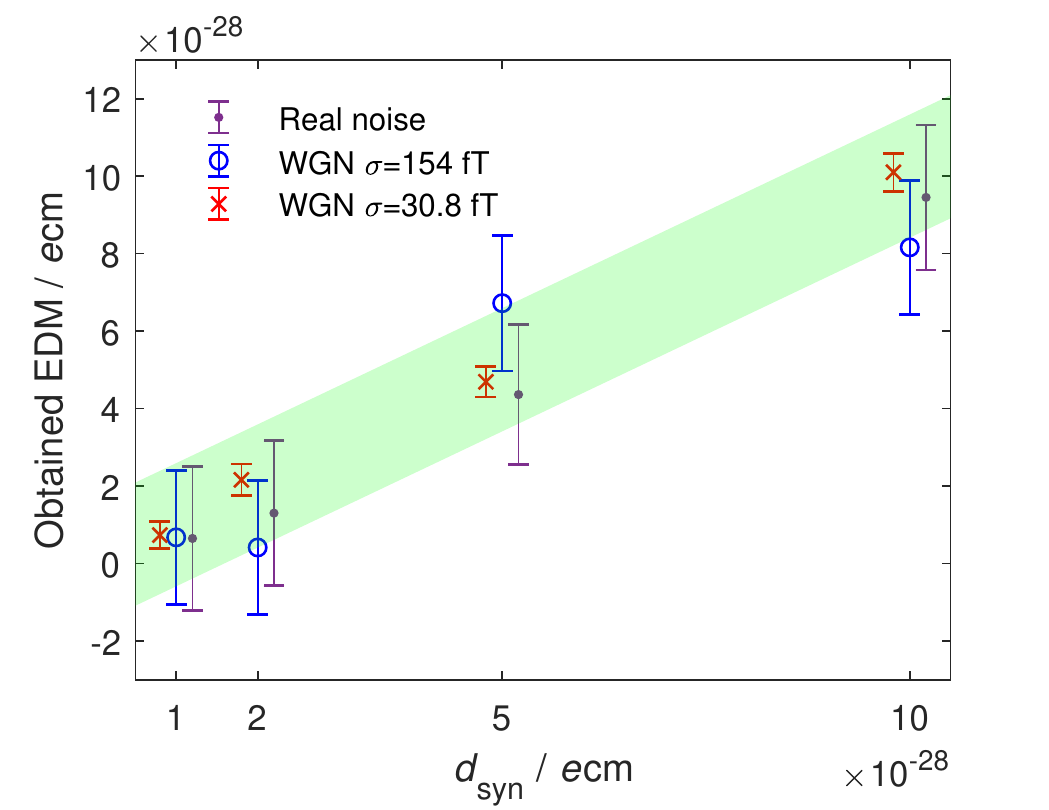}}
    \caption{ \label{fig:MC} The derived EDM values using the synthetic data sets. The $x$ coordinates of the data for real noise and WGN with $\sigma = 30.6 $~fT are shifted with $2 \times 10^{-29}~e~\mathrm{cm}$ and $-2\times 10^{-29}~e~\mathrm{cm}$, respectively. The green shade illustrates the $1\sigma$ confidence interval with the added EDM as the center value and the uncertainty derived from the CRLB for $\sigma = 154 $~fT. }
\end{figure}

The parameters used to generate synthetic data were taken from 18 sub-runs of high sensitivity from the 2018 campaign. The starting amplitude of $^{129}$Xe and $^{3}$He are set to $A_0^{\mathrm{Xe}} =70$~pT, $A_0^{\mathrm{He}}=25$~pT and $T_2^{\mathrm{Xe/He}}=8000$~s. The electric field contains 36 segments of 200~s up to 800~s length, as used in the measurement campaign. The values of other parameters in Eqs.~(\ref{eqn:Phi_Xesyn}) and (\ref{eqn:Phi_Hesyn}) are random and uniformly distributed in the ranges listed in Table~\ref{tab:Syn_para}. Three different kinds of noise were separately added into the synthetic data, including two WGN with $\sigma =154$~fT, the standard deviation of the white noise in the real data, and $\sigma = 154/5=30.8$~fT, as well as real SQUID gradiometer noise. The overall EDM values obtained with the GPF method from the 18 synthetic sub-runs for four synthetic values $d_\mathrm{syn} = (1,2,5,10) \times 10^{-28}~e~\mathrm{cm}$ are plotted in Fig.~\ref{fig:MC}. The averaged overall EDM uncertainty for WGN data with $\sigma =154$~fT is $1.74 \times 10^{-28}~e~\mathrm{cm}$, which is roughly a factor of 5 larger than that obtained from the data with $\sigma = 30.8$~fT and a factor of 1.1 higher than the calculated CRLB for these 18 sub-runs, which is $1.59\times 10^{-28}~e~\mathrm{cm}$. This mainly results from the correlation between the EDM and the parameters of the polynomials in the phase fit. The uncertainty for the real noise is $1.85\times 10^{-28}~e~\mathrm{cm}$, being similar to that for the white noise with $\sigma = 154$ fT. Most of the $1\sigma$ confidence intervals of the derived EDM cover the added EDM values $d_\mathrm{syn}$, showing that the GPF method is capable of accurately obtaining $ d_\mathrm{syn} \geq 1 \times 10^{-28}~e~\mathrm{cm}$ independent of the realistic noise level.
 
\subsection{Overall results}

\subsubsection{Statistical uncertainty}

Applying the GPF method to the same data set of 41 runs (80 sub-runs) as analyzed by the PC method \cite{Sachdeva2019} and using the same channel and analysis parameters, the statistical uncertainty is decreased by a factor of 2.1 from 6.6 $\times 10^{-28}~e~\mathrm{cm}$ to 3.1 $\times 10^{-28}~e~\mathrm{cm}$.

Due to fewer constraints in the GPF method, runs with the number of segments $M \neq 4n$ with $n \in \mathbb{N}$ or having SQUID jumps could be included in the data analysis, leading to a total of 45 runs (87 sub-runs). Furthermore, the segments with zero high voltage are included into the analysis. For the analysis, the block length is $t_\mathrm{b}=5$~s, the threshold of the $F$-test is set to $P_\mathrm{min}=0.6$  (see Appendix~\ref{sec:F_test}) and the minimum order of the polynomial used in the fit is set to 4 in order to adequately describe the comagnetometer phase drift. The average polynomial order used for all sub-runs is 6.4 and the maximum order 13.

\begin{figure}[htpb]
    \centerline{\includegraphics[width=0.6\columnwidth]{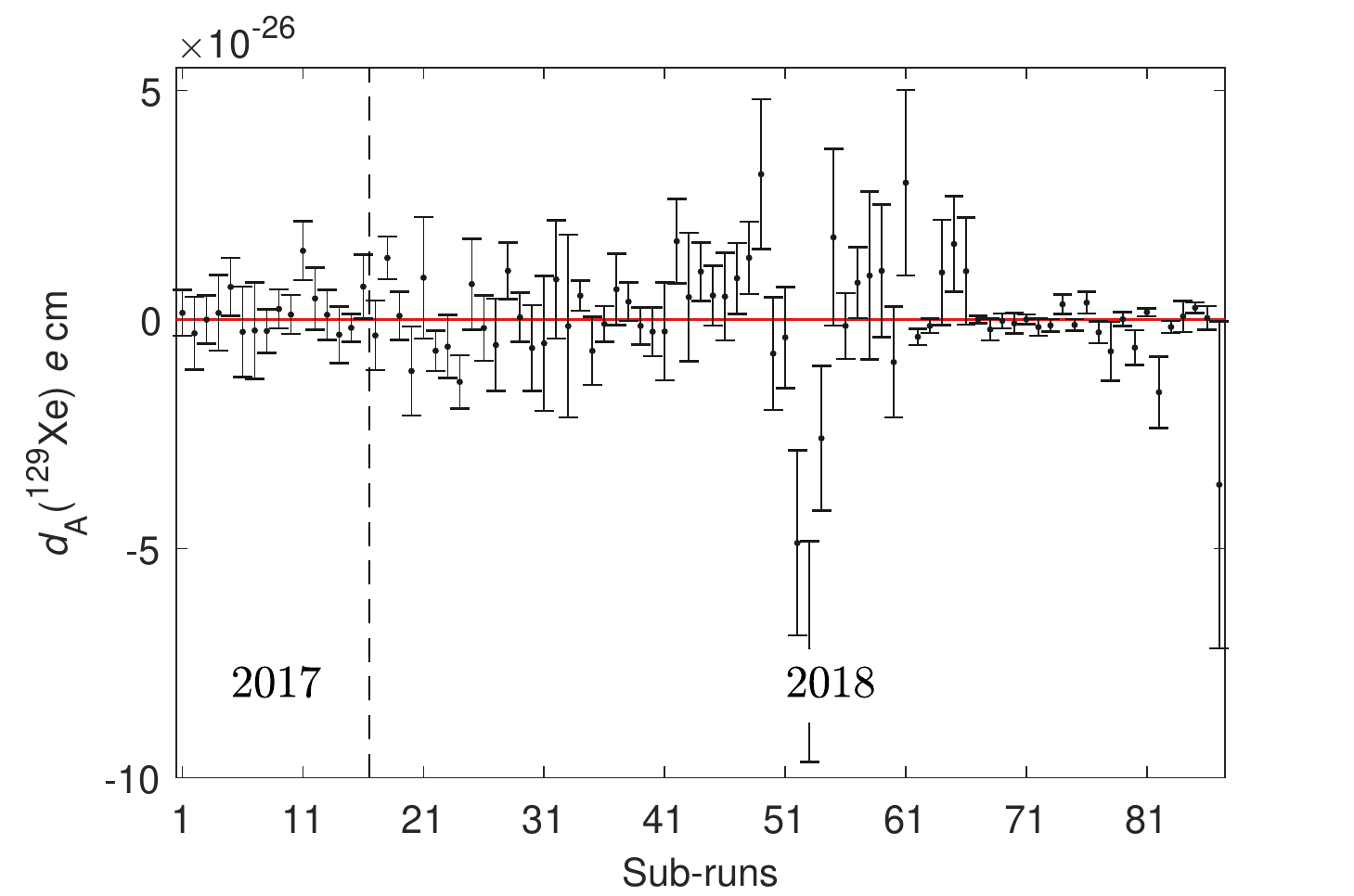}}
    \caption{ \label{fig:Sub_run} EDM results of the 2017 and 2018 campaigns derived with the GPF method by sub-runs. The thin orange bar is the confidence interval of $1\sigma$ around the weighted mean. The reason for a lower uncertainty in the last 20 sub-runs is a change in the experimental parameters as explained in detail in Ref.~\cite{Sachdeva2019}. }
\end{figure}

\begin{figure}[htpb]
    \centerline{\includegraphics[width=0.6\columnwidth]{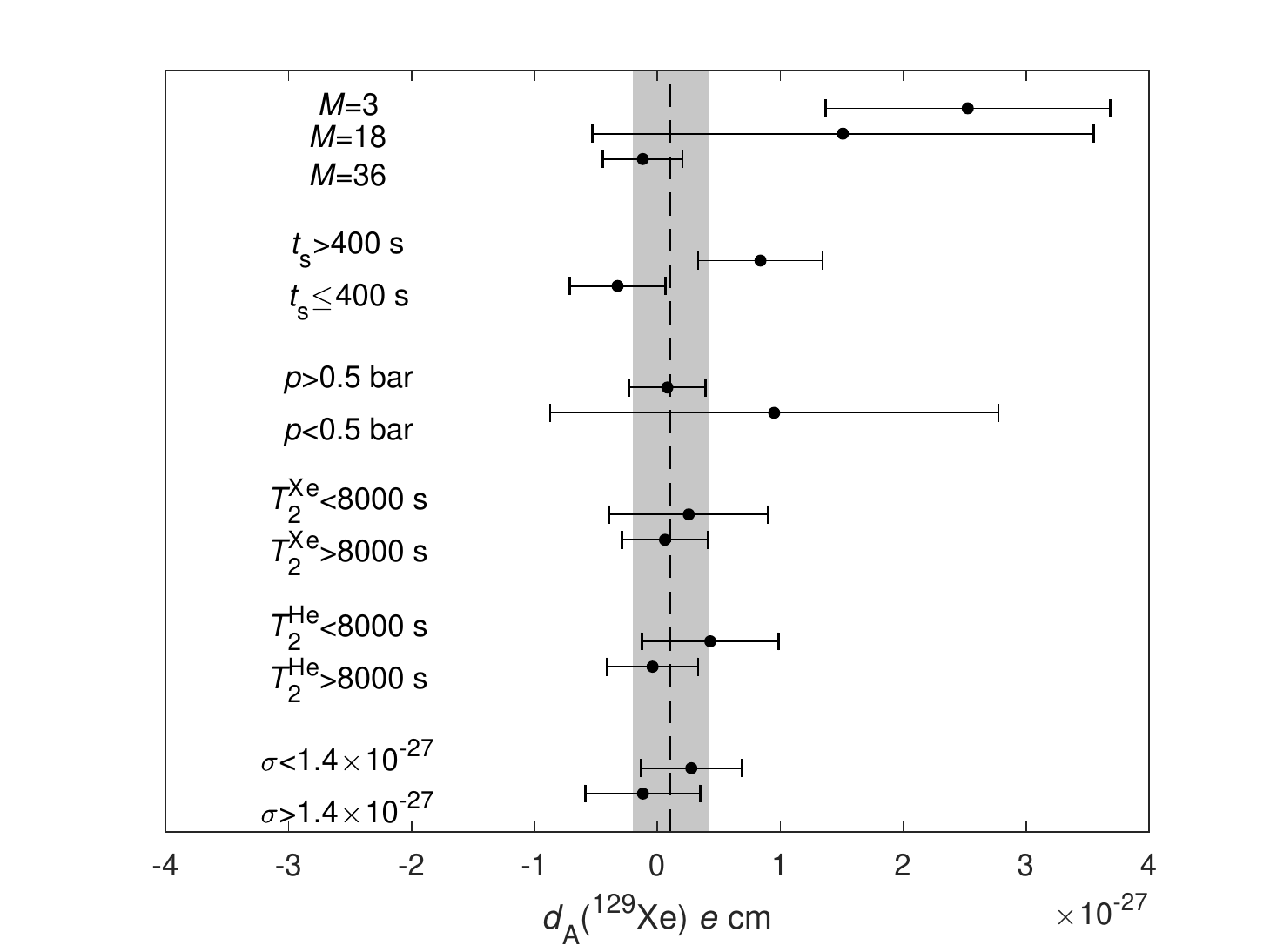}}
    \caption{ \label{fig:Correlation} The EDM results for grouping the data set by number of segments $M$, segment duration $t_s$, gas pressure $p$, $T_2^{Xe}$, $T_2^{He}$, and the statistical uncertainty threshold of  $1.4 \times 10^{-27}~e~\mathrm{cm}$. The dashed line is at $d_\mathrm{A} (^{129}\mathrm{Xe})=1.1 \times 10^{-28}~e~\mathrm{cm}$ and the gray region indicates the confidence interval of $1\sigma$ with the unscaled statistical uncertainty. For clarity of the figure, only a few parameters are plotted here.}
\end{figure}

The overall result using the full data set is $d_\mathrm{A} (^{129}\mathrm{Xe})  = 1.1 \pm 3.1 \times 10^{-28}~e~\mathrm{cm}$  with $\chi^2/\text{\text{dof}}=115.5/86$. As all sub-run measurements were taken with considerable different background noise a $\chi^2/\text{dof}  \geq 1$ can be expected.  According to the PDG guidelines \cite{Beringer2012} we accounted for these random variations by scaling the statistical uncertainty with the factor $\sqrt{\chi^2/\text{dof}}=1.16$  leading to $3.6 \times 10^{-28}~e~\mathrm{cm}$. Bootstrapping \cite{Efron1982} the 87 EDM measurements resulted in an estimate of the statistical uncertainty of $3.14 \times 10^{-28}~e~\mathrm{cm}$. Fig.~\ref{fig:Sub_run} shows the derived EDM results per sub-run. Sorting all EDM measurements into groups based on the experimental parameters, such as the cell geometry, $B_0$ field direction, number and duration of segments and the gas pressure, shows no correlation between the deduced EDM value and these parameters, as can be seen in Fig.~\ref{fig:Correlation}. Furthermore, no correlation between the chosen polynomial order and the derived sub-run EDM values was seen.

\subsubsection{Systematic uncertainty}

The systematic uncertainties of the two experiment campaigns were extensively studied in Ref.~\cite{Sachdeva2019}. We applied the same analysis to the full data set used here and the derived systematic uncertainties are summarized in Table~\ref{tab:GPF_syst}. The correction to the comagnetometer frequency drift of order higher than 1, as it has been done in Ref.~\cite{Sachdeva2019}, becomes obsolete for the GPF method since the model of Eq.~(\ref{eqn:GPF}) considers the higher order drifts implicitly. 

As mentioned above, the GPF uses the full data set, including data during the high voltage rampings. Therefore the charging current can have an impact on the result in two ways. First the charging current could magnetize parts of the experimental equipment, and change the magnetic field seen by the spins. By this mechanism a false EDM may be generated. This effect has been carefully analyzed in Ref.~\cite{Sachdeva2019} and has been adapted for the data set used for GPF (see Charging current in Table~\ref{tab:GPF_syst}). Secondly, the charging currents just as the leakage currents will generate magnetic fields, which are correlated with the electric field direction. This effect is only present during the ramping lasting for a few blocks per segment, ranging from 20 to 160 blocks. The impact of the charging current acting as a leakage current is calculated and turned out to be negligible, relative to the effect of leakage currents as given in Table~\ref{tab:GPF_syst}.

We further looked for the potential effect of the comagnetometer drift and the vibrational noise with Monte-Carlo simulations and did not find observable systematic error, see Appendix~\ref{sec:Non-ideal}.

\begin{table}[h]
    \centering
    \caption{ The systematic uncertainties determined as done in Ref.~\cite{Sachdeva2019} based on the data set used for the GPF method.}
    \begin{tabular}{c c  c }     
    \hline\noalign{\smallskip}
      & 2017 ($e$~cm) & 2018($e$~cm) \\
    \noalign{\smallskip}\hline\noalign{\smallskip}
    Leakage current (incl. impact of $I_\text{Charging}$ during ramping)            &  $ 1.2 \times 10^{-28} $    &  $ 4.4 \times 10^{-31} $ \\
    
    Charging current           &  $ 1.7 \times 10^{-29} $    &  $ 1.2 \times 10^{-29} $  \\
    Cell motion (rotation)     &  $ 4.2 \times 10^{-29} $    &  $ 4.0 \times 10^{-29} $  \\    
    Cell motion (translation)  &  $ 2.6 \times 10^{-28} $    &  $ 1.9 \times 10^{-28} $  \\    
     $|E|^2$ effect            &  $ 1.2 \times 10^{-29} $    &  $ 2.2 \times 10^{-30} $  \\    
     $|E|$ uncertainty         &  $ 9.9 \times 10^{-29} $    &  $ 5.7 \times 10^{-30} $   \\    
     Geometric phase           &  $ \leq 2 \times 10^{-31} $ & $ \leq 1 \times 10^{-29} $ \\   
    \noalign{\smallskip}\hline\noalign{\smallskip}
    \textbf{Total systematic uncertainty}    &  $ 3.07 \times 10^{-28}  $    &  $ 1.95 \times 10^{-28} $ \\  
    \textbf{Scaled statistical uncertainty}         &  $ 15.57 \times 10^{-28} $    &  $ 3.67 \times 10^{-28} $ \\
    \noalign{\smallskip}\hline 
    \end{tabular} 
    \label{tab:GPF_syst}
\end{table}

The overall systematic uncertainty is the weighted average of the systematic uncertainties of the two measurement campaigns 2017 and 2018 using the reciprocal of its statistical variance as weights, yielding $2.0  \times 10 ^{-28}~e~ \mathrm{cm}$. The final result, separating the statistical and systematic uncertainties, is 
\begin{equation}
    \label{eqn:EDM_result}
    d_\mathrm{A} (^{129}\mathrm{Xe})=1.1 \pm 3.6_\mathrm{(stat)} \pm 2.0_\mathrm{(syst)}  \times 10 ^{-28} e \mathrm{cm},
\end{equation}
from which we set an upper limit  $|d_\mathrm{A} (^{129}\mathrm{Xe})| < 8.3 \times 10^{-28}~e~\mathrm{cm}$ at the 95\% C.L. This reanalysis leads to a limit that is a factor of 1.7 smaller compared to the previous result \cite{Sachdeva2019} and a factor of 8.0 compared to the result in 2001 \cite{Rosenberry2001}. 

\section{SUMMARY AND OUTLOOK}

We proposed a global phase fitting method to analyze spin precession data. Applying the GPF method to the data set used in Ref.~\cite{Sachdeva2019} yields a consistent result for $d_\mathrm{A} (^{129}\mathrm{Xe})$ but a two times smaller statistical uncertainty compared to the PC method, as predicted by the theoretical CRLB analysis. Using additional data which had to be discarded for the PC method due to incomplete electric field patterns and optimizing the analysis parameters, the upper limit of the $^{129}$Xe EDM improves by a factor of 1.7 to $|d_\mathrm{A} (^{129}\mathrm{Xe})| < 8.3 \times 10^{-28}~e~\mathrm{cm}$ at the 95\% C.L. This enables $^{129}$Xe to be used as a comagnetometer in future neutron EDM experiments \cite{Masuda2012} with a systematic error contribution down to $ |d_\mathrm{A} (^{129}\mathrm{Xe})|\times \gamma_\text{n}/\gamma_\text{Xe}=2.1 \times 10^{-27}~e~\mathrm{cm}$. Our GPF method relieves the demands on the physical model describing the comagnetometer frequency drift and could be generally used in similar spin precession experiments, such as the Lorentz-invariance test. By optimizing the experimental parameters for the GPF method (see Appendix~\ref{sec:future}), the upper limit for $d_\mathrm{A} (^{129}\mathrm{Xe})$ could be reduced even further, as planned for an upcoming EDM campaign with optimized high voltage pattern.

\section*{Acknowledgements}
We acknowledge the support of the Core Facility ‘Metrology of Ultra-Low Magnetic Fields’ at Physikalisch-Technische Bundesanstalt which receives funding from the Deutsche Forschungsgemeinschaft (DFG KO 5321/3-1 and TR 408/11-1). This work was supported by Deutsche Forschungsgemeinschaft grants TR408/12 and FA1456/1-1. T. Liu acknowledges the support from the Funds for International Cooperation and Exchange of the National Natural Science Foundation of China (grant number 51861135308). We gratefully thank Dr. U. Steinhoff and Dr. G. W\"ubbeler for fruitful discussion.

\section*{Authors contributions}
All the authors collaborated in the same way in the ideas, development and writing of the manuscript.

\appendix
\section{ EXPERIMENT DEPENDENT FACTORS}
\label{sec:Non-ideal}
Here we applied Monte-Carlo simulation to investigate the impact of two non-ideal factors on the derived EDM result, namely the vibrational noise and the comagnetometer phase drift. 

\subsection{Vibrational noise }
\label{subsec:Vibra}
The effect of the real measurement noise (e.g. vibrational noise) on the estimated phase is quantitatively analyzed. Here, synthetic data is generated, using a single sinusoidal function with a constant amplitude $A = 30$~pT, with a length of 10000~s yielding 2000 blocks. Furthermore, white noise with $\sigma =154$~fT (the standard deviation of the white noise in real gradiometer data) generated with MATLAB, or real noise (from the exemplary sub-run with a total noise power of 0.4~pT and the precession signals filtered out) were added separately to the synthetic data. The error of the fitted phase for block $i$ is defined as $\epsilon_i =\phi_\mathrm{fit,i}-\phi_\mathrm{real,i}$. Here $\phi_\mathrm{real,i}$ is known and $\phi_\mathrm{fit,i}$ is obtained from the fit to block $i$. 
\begin{figure}[htpb]
    \centerline{\includegraphics[width=0.6\columnwidth]{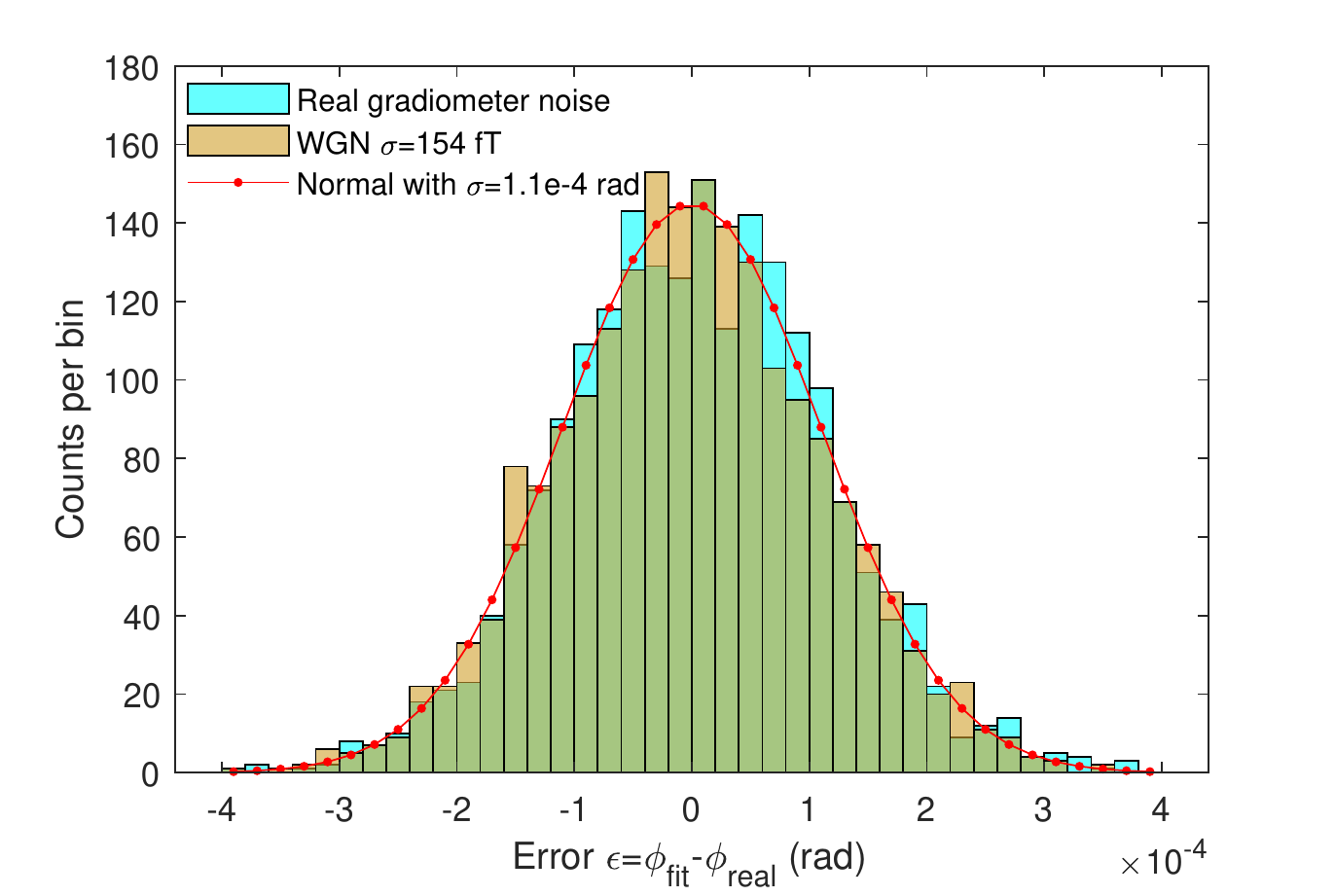}}
    \caption{ \label{fig:Vib} Histograms of the phase error from the synthetic data with real gradiometer noise (light blue) and white noise (yellow). The data lasts for 10 000 s and consists of 200 blocks. }
\end{figure}

The histograms of $\epsilon_i$ for these two synthetic data sets are plotted in Fig.~\ref{fig:Vib}. The error for the white noise data is in good agreement with the normal distribution with $\sigma = 1.11\times 10^{-4} $ rad, which is close to  $1.08\times 10^{-4} $ rad, the CRLB on the phase estimator in Eq.~(\ref{eqn:CRLB_phase}). The error for the real noise data also satisfies the Gaussian distribution, with a similar result as the white noise data. This implies that the vibrational noise did not cause evident additional phase deviation, although the standard deviation of the vibrational noise is around 7 times bigger than the white noise. As evident in Fig.~\ref{fig:Vib}, the vibrational noise does not cause an observable systematic error on the derived phase. 

\begin{figure*}[htbp]
    \centerline{\includegraphics[width=1\columnwidth]{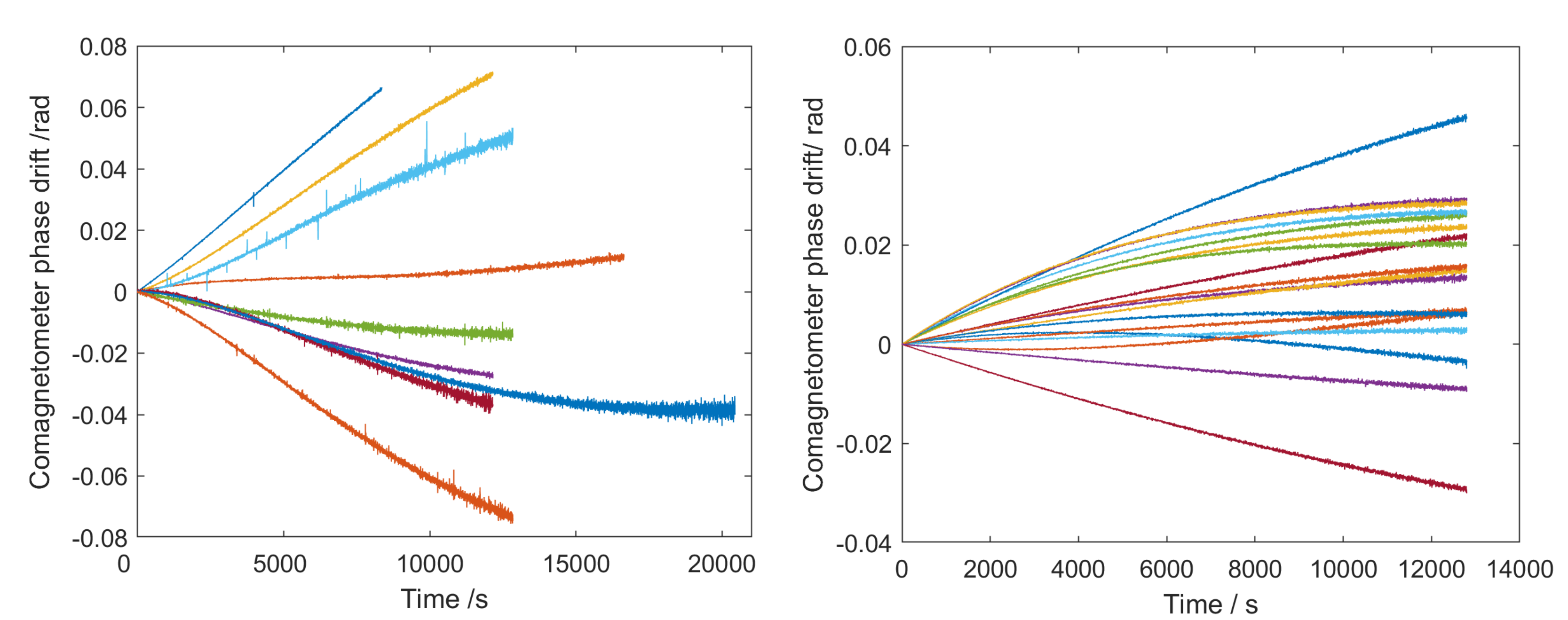}}
    \caption{ \label{fig:Drift}  Left: Measured comagnetometer phase drift of 9 sub-runs reduced by the linear deterministic term stemming from the Earth's rotation and chemical shift. The green curve is the result of the exemplary sub-run used in Sec.~\ref{sec:GPF}. Right: synthetic data of 18 sub-runs. }
\end{figure*}

\subsection{ Comagnetometer phase drift }
\label{subsec:phase_drift}
The analyzed comagnetometer phase drift  $\Phi_{\mathrm{co}}^k$ for 9 sub-runs with high sensitivities on $^{129}$Xe EDM are plotted in the left panel of Fig.~\ref{fig:Drift}. Note that the linear drifts due to Earth’s rotation and chemical shift were subtracted by using the deterministic equations \cite{Gemmel2010}. The synthetic comagnetometer phase drifts generated with two exponential functions for 18 random sub-runs are plotted in the right panel of Fig.~\ref{fig:Drift}, showing a similar behavior as the experimentally obtained ones. The parameter ranges are listed in Table~\ref{tab:Syn_para}.

\begin{figure}[htpb]
    \centerline{\includegraphics[width=0.6\columnwidth]{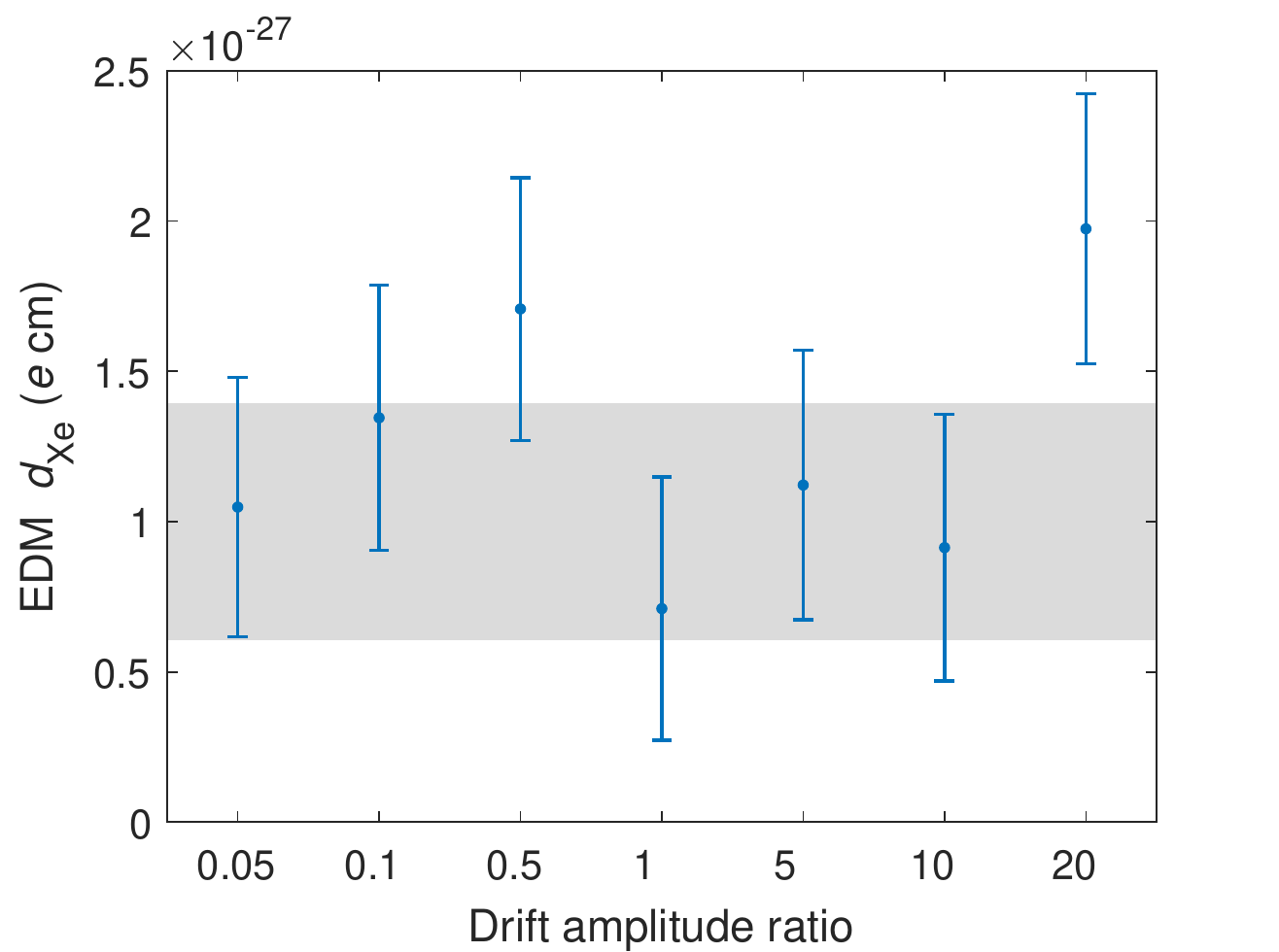}}
    \caption{ \label{fig:Drift_result} The derived EDM value as a function of the scale ratio of the drift amplitude to the observed drift in two campaigns. The uncertainty is for $1\sigma$. Each result is an average of 40 sub-runs lasting 12800 s and with 32 high voltage segments. The gray bar indicates the $1\sigma$ confidence interval with the added EDM as the center value and the uncertainty derived from the CRLB.}
\end{figure}

To investigate the potential systematic effect caused by the comagnetometer phase drift, we altered the drift amplitude $u^\mathrm{Xe}$ and $u^\mathrm{He}$ in Eqs.~(\ref{eqn:Phi_Xesyn}) and (\ref{eqn:Phi_Hesyn}) in the synthetic phase data. Fig.~\ref{fig:Drift_result} shows the derived EDM values as a function of the scale ratio of the drift amplitude. No distinct correlation between the obtained EDM value and the drift amplitude could be observed. Therefore, we did not assign a model dependent uncertainty for the comagnetometer drift when applying the GPF method. 

\section{ THE ${F}$-TEST THRESHOLD}
\label{sec:F_test}
The $F$-test threshold $P_\mathrm{min}$ affects the polynomial order used in the GPF method, as listed in Table~\ref{tab:GPF_F_test}.  The EDM values for various $P_\mathrm{min}$ are overlapped within the $1\sigma$ statistical uncertainty and are all consistent with zero. Additionally, the upper limit of the $^{129}$Xe EDM is almost insensitive to the threshold. We have chosen 0.6 as $F$-test threshold yielding the highest upper bound.

\begin{table*}[h]
    \centering
    \caption{ The overall EDM results with various $F$-test threshold $P_\mathrm{min}$. }
    \begin{tabular}{c c c c c c c}
    \hline\noalign{\smallskip}
     $P_{\mathrm{min}}$ & Average order & EDM  & Uncertainty & Reduced $\chi^2$ & $P$-value & Upper limit (95\% C.L.)  \\
       & & ($10^{-28} e$~cm) &  ($10^{-28} e$~cm) &   &  & ($10^{-28} e$~cm)  \\
    \noalign{\smallskip}\hline\noalign{\smallskip}
    0.4  & 8.2  &  0.08  &  3.22 & 1.32 & 0.03  & 8.2  \\    
    0.5  & 7.3  & -0.36   &  3.20 & 1.24 &0.06  & 8.0  \\
    0.6  & 6.4  &  1.06   &  3.08 & 1.34 &0.02  & 8.3  \\
    0.7  & 6.0  & -0.07  &  3.06 & 1.26 & 0.05  & 7.8  \\    
    0.8  & 5.5  & -0.87  &  3.05 & 1.31 & 0.03  & 8.1  \\    
    
    \noalign{\smallskip}\hline
    \end{tabular}
    \label{tab:GPF_F_test}
\end{table*}

\section{ DESIGN OF EXPERIMENTAL PARAMETERS}
\label{sec:future}
\begin{figure}[htpb]
    \centerline{\includegraphics[width=0.6\columnwidth]{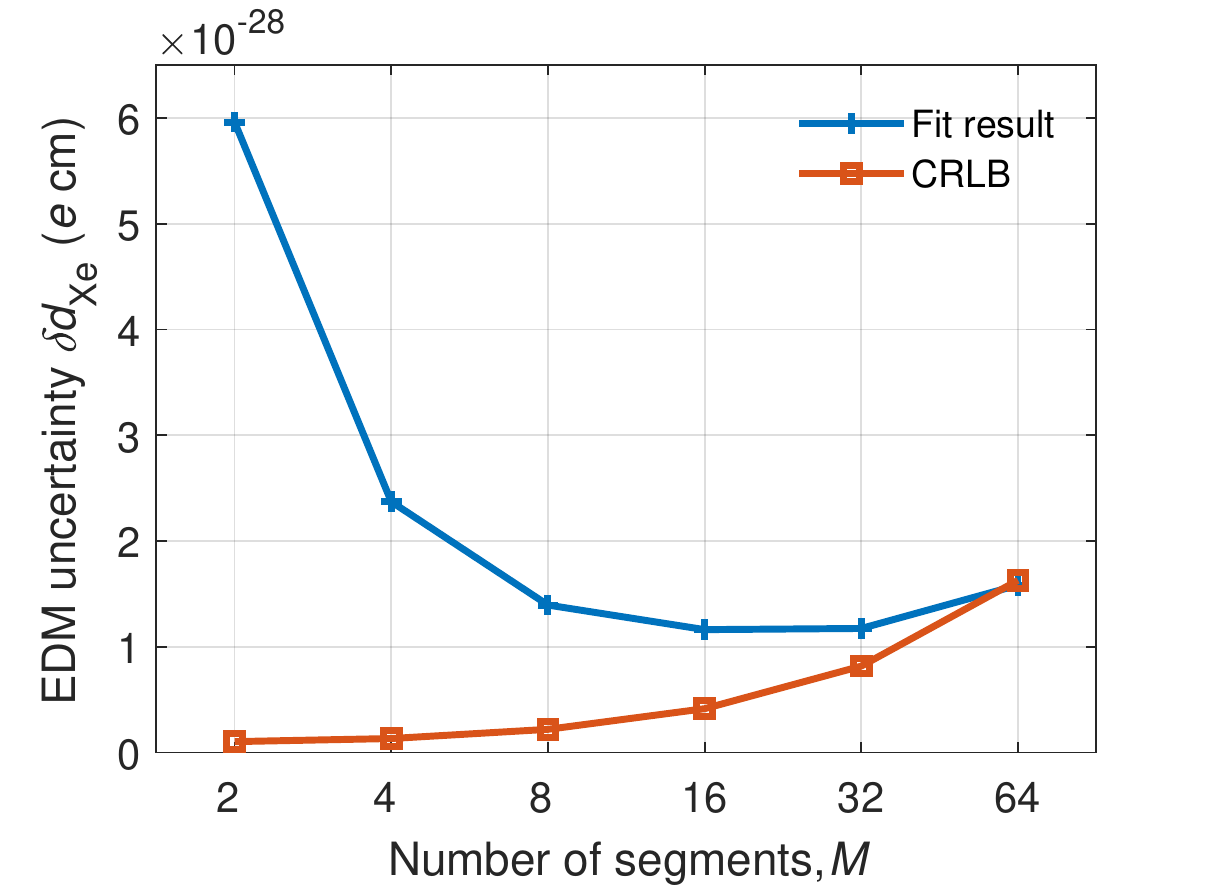}}
    \caption{ \label{fig:Future} The averaged EDM value over 100 random simulated runs. The reduced chi-square value was normalized to 1 for each result. The solid lines are guides for the eye.}
\end{figure}

The number of segments $M$ in one sub-run has a significant impact on the estimation uncertainty derived by the GPF method. According to the ideal CRLB, a smaller number of segments results in a lower uncertainty, shown as the red line in Fig.~\ref{fig:Future}. To search for the optimum segment number, we used the synthetic comagnetometer phase data with added white Gaussian noise. The phase uncertainty increases with time and starts with 0.1~mrad. The time constants $T_2$ for $^{129}$Xe atoms and $^{3}$He atoms are a random number ranged from 8000~s to 9000~s. The total measurement time length is fixed to 38400~s, while $M$ is varied from 2 to 64. The averaged EDM value over 100 runs for each $M$ are plotted as the blue crosses. The fit uncertainty is larger than the ideal CRLB due to the correlation between the EDM function and the phase drift. The gap is reduced with the increase of $M$, since the orthogonality condition is satisfied better. A relatively flat optimum is found around $M$~=~16.  Note that this optimum value also depends on the total measurement time. A sub-run with longer measurement time calls for a higher number of segments, hence the optimum number for $T = 6400$~s and $T = 64000$~s is 8 and 64, respectively. The improved understanding of the comagnetometer frequency drift behavior may reduce the requirement on the segment number, thus significantly increasing the measurement sensitivity. 

\printbibliography

@PREAMBLE{
 "\providecommand{\noopsort}[1]{}" 
 # "\providecommand{\singleletter}[1]{#1}%" 
}

@article{Sakharov1967,
author = {Sakharov, Andrei D.},
issn = {0038-5670},
journal = {JETP Letters},
month = {may},
number = {5},
pages = {24--27},
title = {{Violation of CP in variance, C asymmetry, and baryon asymmetry of the universe}},
volume = {5},
year = {1967}
}

@article{Canetti2012,
author = {Canetti, Laurent and Drewes, Marco and Shaposhnikov, Mikhail},
doi = {10.1088/1367-2630/14/9/095012},
issn = {1367-2630},
journal = {New Journal of Physics},
month = {sep},
number = {9},
pages = {095012},
title = {{Matter and antimatter in the universe}},
url = {http://stacks.iop.org/1367-2630/14/i=9/a=095012?key=crossref.f53af890281fe28cd0946a77548a9b63},
volume = {14},
year = {2012}
}

@article{Dine2003,
author = {Dine, Michael and Kusenko, Alexander},
doi = {10.1103/RevModPhys.76.1},
issn = {0034-6861},
journal = {Reviews of Modern Physics},
month = {dec},
number = {1},
pages = {1--30},
title = {{Origin of the matter-antimatter asymmetry}},
url = {https://link.aps.org/doi/10.1103/RevModPhys.76.1},
volume = {76},
year = {2003}
}

@article{Smith1957,
author = {Smith, J. H. and Purcell, E. M. and Ramsey, N. F.},
doi = {10.1103/PhysRev.108.120},
issn = {0031-899X},
journal = {Physical Review},
month = {oct},
number = {1},
pages = {120--122},
title = {{Experimental Limit to the Electric Dipole Moment of the Neutron}},
url = {https://doi.org/10.1103/PhysRev.108.120 https://link.aps.org/doi/10.1103/PhysRev.108.120},
volume = {108},
year = {1957}
}

@article{Chupp2019,
archivePrefix = {arXiv},
arxivId = {1710.02504v1},
author = {Chupp, T. E. and Fierlinger, P and Ramsey-Musolf, M. J. and Singh, J. T.},
doi = {10.1103/RevModPhys.91.015001},
eprint = {1710.02504v1},
issn = {0034-6861},
journal = {Reviews of Modern Physics},
month = {jan},
number = {1},
pages = {015001},
title = {{Electric dipole moments of atoms, molecules, nuclei, and particles}},
url = {https://link.aps.org/doi/10.1103/RevModPhys.91.015001},
volume = {91},
year = {2019}
}

@article{Terrano2019,
archivePrefix = {arXiv},
arxivId = {1807.11119},
author = {Terrano, W. A. and Meinel, Jonas and Sachdeva, Natasha and Chupp, T. E. and Degenkolb, Skyler and Fierlinger, Peter and Kuchler, Florian and Singh, Jaideep T.},
doi = {10.1103/PhysRevA.100.012502},
eprint = {1807.11119},
issn = {2469-9926},
journal = {Physical Review A},
month = {jul},
number = {1},
pages = {012502},
title = {{Frequency shifts in noble-gas comagnetometers}},
url = {http://arxiv.org/abs/1807.11119 https://link.aps.org/doi/10.1103/PhysRevA.100.012502},
volume = {100},
year = {2019}
}

@article{Limes2019,
archivePrefix = {arXiv},
arxivId = {arXiv:1805.11578v2},
author = {Limes, M E and Dural, N and Romalis, M V and Foley, E. L. and Kornack, T. W. and Nelson, A. and Grisham, L. R. and Vaara, J.},
doi = {10.1103/PhysRevA.100.010501},
eprint = {arXiv:1805.11578v2},
issn = {2469-9926},
journal = {Physical Review A},
month = {jul},
number = {1},
pages = {010501},
title = {{Dipolar and scalar $^{3}$He - $^{129}$Xe frequency shifts in stemless cells}},
url = {https://link.aps.org/doi/10.1103/PhysRevA.100.010501},
volume = {100},
year = {2019}
}

@article{Sachdeva2019,
archivePrefix = {arXiv},
arxivId = {1902.02864},
author = {Sachdeva, N. and Fan, I. and Babcock, E. and Burghoff, M. and Chupp, T. E. and Degenkolb, S. and Fierlinger, P. and Haude, S. and Kraegeloh, E. and Kilian, W. and Knappe-Gr{\"{u}}neberg, S. and Kuchler, F. and Liu, T. and Marino, M. and Meinel, J. and Rolfs, K. and Salhi, Z. and Schnabel, A. and Singh, J. T. and Stuiber, S. and Terrano, W. A. and Trahms, L. and Voigt, J.},
doi = {10.1103/PhysRevLett.123.143003},
eprint = {1902.02864},
issn = {0031-9007},
journal = {Physical Review Letters},
month = {oct},
number = {14},
pages = {143003},
title = {{New Limit on the Permanent Electric Dipole Moment of $^{129}$Xe Using $^{3}$He Comagnetometry and SQUID Detection}},
url = {http://arxiv.org/abs/1902.02864 https://link.aps.org/doi/10.1103/PhysRevLett.123.143003},
volume = {123},
year = {2019}
}

@phdthesis{Sachdeva2019a,
author = {Sachdeva, Natasha},
isbn = {0000000193054},
school = {The University of Michigan},
title = {{A Measurement of the Permanent Electric Dipole Moment of $^{129}$Xe}},
year = {2019}
}

@article{Cairncross2019,
author = {Cairncross, William B. and Ye, Jun},
doi = {10.1038/s42254-019-0080-0},
issn = {2522-5820},
journal = {Nature Reviews Physics},
number = {8},
pages = {510--521},
publisher = {Springer US},
title = {{Atoms and molecules in the search for time-reversal symmetry violation}},
url = {http://dx.doi.org/10.1038/s42254-019-0080-0},
volume = {1},
year = {2019}
}

@article{Graner2016,
author = {Graner, B. and Chen, Y. and Lindahl, E. G. and Heckel, B. R.},
doi = {10.1103/PhysRevLett.116.161601},
issn = {10797114},
journal = {Physical Review Letters},
number = {16},
pages = {1--5},
title = {{Reduced Limit on the Permanent Electric Dipole Moment of $^{199}$Hg}},
volume = {116},
year = {2016}
}

@article{Bishof2016,
author = {Bishof, Michael and Parker, Richard H. and Bailey, Kevin G. and Greene, John P. and Holt, Roy J. and Kalita, Mukut R. and Korsch, Wolfgang and Lemke, Nathan D. and Lu, Zheng Tian and Mueller, Peter and O'Connor, Thomas P. and Singh, Jaideep T. and Dietrich, Matthew R.},
doi = {10.1103/PhysRevC.94.025501},
issn = {24699993},
journal = {Physical Review C},
number = {2},
pages = {1--17},
title = {{Improved limit on the $^{225}$Ra electric dipole moment}},
volume = {94},
year = {2016}
}

@article{Sakurai2019,
author = {Sakurai, Akitada and Sahoo, B K and Asahi, K and Das, B P},
doi = {10.1103/PhysRevA.100.020502},
issn = {2469-9926},
journal = {Physical Review A},
month = {aug},
number = {2},
pages = {020502},
publisher = {American Physical Society},
title = {{Relativistic many-body theory of the electric dipole moment of $^{129}$Xe and its implications for probing new physics beyond the standard model}},
url = {https://link.aps.org/doi/10.1103/PhysRevA.100.020502},
volume = {100},
year = {2019}
}

@article{Abel2020,
archivePrefix = {arXiv},
arxivId = {2001.11966},
author = {Abel, C and Afach, S and Ayres, N. J. and Baker, C. A. and Ban, G and Bison, G and Bodek, K and Bondar, V and Burghoff, M and Chanel, E and Chowdhuri, Z and Chiu, P. J. and Clement, B and Crawford, C. B. and Daum, M and Emmenegger, S and Ferraris-Bouchez, L. and Fertl, M and Flaux, P and Franke, B and Fratangelo, A and Geltenbort, P and Green, K and Griffith, W. C. and {Van Der Grinten}, M. and Gruji{\'{c}}, Z. D. and Harris, P. G. and Hayen, L. and Heil, W. and Henneck, R. and H{\'{e}}laine, V. and Hild, N. and Hodge, Z. and Horras, M. and Iaydjiev, P. and Ivanov, S. N. and Kasprzak, M. and Kermaidic, Y. and Kirch, K. and Knecht, A. and Knowles, P. and Koch, H. C. and Koss, P. A. and Komposch, S. and Kozela, A. and Kraft, A. and Krempel, J. and Ku{\'{z}}niak, M. and Lauss, B. and Lefort, T. and Lemi{\`{e}}re, Y. and Leredde, A. and Mohanmurthy, P. and Mtchedlishvili, A. and Musgrave, M. and Naviliat-Cuncic, O. and Pais, D. and Piegsa, F. M. and Pierre, E. and Pignol, G. and Plonka-Spehr, C. and Prashanth, P. N. and Qu{\'{e}}m{\'{e}}ner, G. and Rawlik, M. and Rebreyend, D. and Rien{\"{a}}cker, I. and Ries, D. and Roccia, S. and Rogel, G. and Rozpedzik, D. and Schnabel, A. and Schmidt-Wellenburg, P. and Severijns, N. and Shiers, D. and {Tavakoli Dinani}, R. and Thorne, J. A. and Virot, R. and Voigt, J. and Weis, A. and Wursten, E. and Wyszynski, G. and Zejma, J. and Zenner, J. and Zsigmond, G.},
doi = {10.1103/PhysRevLett.124.081803},
eprint = {2001.11966},
issn = {10797114},
journal = {Physical Review Letters},
month = {feb},
number = {8},
pages = {081803},
publisher = {American Physical Society},
title = {{Measurement of the Permanent Electric Dipole Moment of the Neutron}},
url = {https://doi.org/10.1103/PhysRevLett.124.081803 https://link.aps.org/doi/10.1103/PhysRevLett.124.081803},
volume = {124},
year = {2020}
}

@article{Flambaum2020,
archivePrefix = {arXiv},
arxivId = {1912.13129},
author = {Flambaum, V. V. and Pospelov, M. and Ritz, A. and Stadnik, Y. V.},
doi = {10.1103/PhysRevD.102.035001},
eprint = {1912.13129},
issn = {2470-0010},
journal = {Physical Review D},
month = {aug},
number = {3},
pages = {035001},
publisher = {American Physical Society},
title = {{Sensitivity of EDM experiments in paramagnetic atoms and molecules to hadronic CP violation}},
url = {http://arxiv.org/abs/1912.13129 https://link.aps.org/doi/10.1103/PhysRevD.102.035001},
volume = {102},
year = {2020}
}

@article{Baker2006,
archivePrefix = {arXiv},
arxivId = {0704.1354},
author = {Baker, C. A. and Doyle, D. D. and Geltenbort, P. and Green, K. and van der Grinten, M. G. D. and Harris, P. G. and Iaydjiev, P. and Ivanov, S. N. and May, D. J. R. and Pendlebury, J. M. and Richardson, J. D. and Shiers, D. and Smith, K. F.},
doi = {10.1103/PhysRevLett.97.131801},
eprint = {0704.1354},
issn = {0031-9007},
journal = {Physical Review Letters},
month = {sep},
number = {13},
pages = {131801},
title = {{Improved Experimental Limit on the Electric Dipole Moment of the Neutron}},
url = {http://arxiv.org/abs/0704.1354{\%}0Ahttp://dx.doi.org/10.1103/PhysRevLett.98.149102 https://link.aps.org/doi/10.1103/PhysRevLett.97.131801},
volume = {97},
year = {2006}
}

@article{Cho1989,
author = {Cho, D. and Sangster, K. and Hinds, E. A.},
journal = {Physical Review Letters},
doi = {10.1103/PhysRevLett.63.2559},
issn = {00319007},
number = {23},
pages = {2559--2562},
title = {{Tenfold improvement of limits on T violation in thallium fluoride}},
volume = {63},
year = {1989}
}

@article{Rosenberry2001,
author = {Rosenberry, M. A. and Chupp, T. E.},
doi = {10.1103/PhysRevLett.86.22},
issn = {0031-9007},
journal = {Physical Review Letters},
month = {jan},
number = {1},
pages = {22--25},
title = {{Atomic Electric Dipole Moment Measurement Using Spin Exchange Pumped Masers of  $^{129}$Xe and $^3$He}},
url = {https://link.aps.org/doi/10.1103/PhysRevLett.86.22},
volume = {86},
year = {2001}
}

@article{Allmendinger2014a,
author = {Allmendinger, F. and Schmidt, U. and Heil, W. and Karpuk, S. and Scharth, A. and Sobolev, Yu and Tullney, K.},
doi = {10.1103/PhysRevLett.113.188902},
issn = {0031-9007},
journal = {Physical Review Letters},
month = {oct},
number = {18},
pages = {188902},
title = {{Allmendinger et al. Reply:}},
url = {https://link.aps.org/doi/10.1103/PhysRevLett.113.188902},
volume = {113},
year = {2014}
}

@article{Romalis2014,
author = {Romalis, Michael V. and Sheng, Dong and Saam, Brian and Walker, Thad G.},
doi = {10.1103/PhysRevLett.113.188901},
issn = {0031-9007},
journal = {Physical Review Letters},
month = {oct},
number = {18},
pages = {188901},
title = {{Comment on New Limit on Lorentz-Invariance- and CPT-Violating Neutron Spin Interactions Using a Free-Spin-Precession $^3$He-$^{129}$Xe Comagnetometer}},
url = {https://link.aps.org/doi/10.1103/PhysRevLett.113.188901},
volume = {113},
year = {2014}
}

@article{Thrasher2019,
archivePrefix = {arXiv},
arxivId = {1910.02156},
author = {Thrasher, D. A. and Sorensen, S. S. and Weber, J. and Bulatowicz, M. and Korver, A. and Larsen, M. and Walker, T. G.},
doi = {10.1103/PhysRevA.100.061403},
eprint = {1910.02156},
issn = {2469-9926},
journal = {Physical Review A},
month = {dec},
number = {6},
pages = {061403},
title = {{Continuous comagnetometry using transversely polarized Xe isotopes}},
url = {http://arxiv.org/abs/1910.02156 https://link.aps.org/doi/10.1103/PhysRevA.100.061403},
volume = {100},
year = {2019}
}

@article{Flambaum2012,
author = {Flambaum, V. V. and Kozlov, A.},
doi = {10.1103/PhysRevA.85.022505},
issn = {10502947},
journal = {Physical Review A},
number = {2},
pages = {1--7},
title = {{Extension of the Schiff theorem to ions and molecules}},
volume = {85},
year = {2012}
}

@article{Tullney2013,
archivePrefix = {arXiv},
arxivId = {1303.6612},
author = {Tullney, K. and Allmendinger, F. and Burghoff, M. and Heil, W. and Karpuk, S. and Kilian, W. and Knappe-Gr{\"{u}}neberg, S. and M{\"{u}}ller, W. and Schmidt, U and Schnabel, A and Seifert, F and Sobolev, Yu and Trahms, L},
doi = {10.1103/PhysRevLett.111.100801},
eprint = {1303.6612},
issn = {00319007},
journal = {Physical Review Letters},
month = {sep},
number = {10},
pages = {100801},
title = {{Constraints on spin-dependent short-range interaction between nucleons}},
url = {https://link.aps.org/doi/10.1103/PhysRevLett.111.100801},
volume = {111},
year = {2013}
}

@article{Golub2003,
author = {Golub, Gene and Pereyra, Victor},
doi = {10.1088/0266-5611/19/2/201},
issn = {0266-5611},
journal = {Inverse Problems},
month = {apr},
number = {2},
pages = {R1--R26},
title = {{Separable nonlinear least squares: the variable projection method and its applications}},
url = {http://stacks.iop.org/0266-5611/19/i=2/a=201?key=crossref.b35e35c4a03d23f59856db3017f18479 https://iopscience.iop.org/article/10.1088/0266-5611/19/2/201},
volume = {19},
year = {2003}
}

@inproceedings{Allan1981,
address = {Philadelphia, Pennsylvania, USA,},
author = {Allan, D.W. and Barnes, J.A.},
doi = {10.1109/freq.1981.200514},
editor = {{Thirty Fifth Annual Frequency Control Symposium}},
isbn = {9788578110796},
issn = {1098-6596},
pages = {470--475},
pmid = {25246403},
publisher = {IEEE},
title = {{A Modified "Allan Variance" with Increased Oscillator Characterization Ability}},
url = {http://ieeexplore.ieee.org/document/1537454/},
year = {1981}
}

@book{KAY1993,
address = {New Jersey},
author = {KAY, STEVEN M.},
isbn = {9780133457117},
pages = {57},
publisher = {Prentice Hall PTR},
title = {{Fundamentals of Statistical Signal Processing, Volume I: Estimation Theory.}},
year = {1993}
}

@book{Efron1982,
address = {Philadelphia},
author = {Efron, Bradley},
publisher = {Society for Industrial and Applied Mathematics},
title = {{The jackknife, the bootstrap, and other resampling plans}},
year = {1982}
}

@book{Refaat2009,
address = {South Carolina},
author = {Refaat El Attar},
publisher = {CreateSpace Independent Publishing Platform},
title = {{Legendre Polynomials And Functions}},
year = {2009}
}

@article{Masuda2012,
author = {Masuda, Yasuhiro and Asahi, Koichiro and Hatanaka, Kichiji and Jeong, Sun Chan and Kawasaki, Shinsuke and Matsumiya, Ryohei and Matsuta, Kensaku and Mihara, Mototsugu and Watanabe, Yutaka},
doi = {10.1016/j.physleta.2012.02.056},
issn = {03759601},
journal = {Physics Letters, Section A: General, Atomic and Solid State Physics},
number = {16},
pages = {1347--1351},
publisher = {Elsevier B.V.},
title = {{Neutron electric dipole moment measurement with a buffer gas comagnetometer}},
url = {http://dx.doi.org/10.1016/j.physleta.2012.02.056},
volume = {376},
year = {2012}
}

@article{Allmendinger2019,
author = {Allmendinger, F. and Engin, I. and Heil, W. and Karpuk, S. and Krause, H.-J. and Niederl{\"{a}}nder, B. and Offenh{\"{a}}usser, A. and Repetto, M. and Schmidt, U. and Zimmer, S.},
doi = {10.1103/PhysRevA.100.022505},
issn = {2469-9926},
journal = {Physical Review A},
number = {2},
pages = {022505},
title = {{Measurement of the permanent electric dipole moment of the $^{129}$Xe atom}},
url = {https://link.aps.org/doi/10.1103/PhysRevA.100.022505},
volume = {100},
year = {2019}
}

@article{Golub1994,
author = {Golub, R. and S. K. Lamoreaux.},
doi = {10.1016/0370-1573(94)90084-1},
issn = {03701573},
journal = {Physics Reports},
month = {feb},
number = {1},
pages = {1--62},
title = {{Neutron electric-dipole moment, ultracold neutrons and polarized 3He}},
url = {https://linkinghub.elsevier.com/retrieve/pii/0370157394900841},
volume = {237},
year = {1994}
}

@article{Gemmel2010,
author = {Gemmel, C. and Heil, W. and Karpuk, S. and Lenz, K. and Ludwig, Ch and Sobolev, Yu and Tullney, K. and Burghoff, M. and Kilian, W. and Knappe-Gr{\"{u}}neberg, S. and M{\"{u}}ller, W. and Schnabel, A. and Seifert, F. and Trahms, L. and Baeler, St},
doi = {10.1140/epjd/e2010-00044-5},
issn = {14346060},
journal = {European Physical Journal D},
number = {3},
pages = {303--320},
title = {{Ultra-sensitive magnetometry based on free precession of nuclear spins}},
volume = {57},
year = {2010}
}

@article{Allmendinger2014,
archivePrefix = {arXiv},
arxivId = {1312.3225},
author = {Allmendinger, F. and Heil, W. and Karpuk, S. and Kilian, W. and Scharth, A. and Schmidt, U. and Schnabel, A. and Sobolev, Yu and Tullney, K.},
doi = {10.1103/PhysRevLett.112.110801},
eprint = {1312.3225},
issn = {00319007},
journal = {Physical Review Letters},
number = {11},
pages = {1--5},
pmid = {27610834},
title = {{New Limit on Lorentz-Invariance- and CPT-Violating Neutron Spin Interactions Using a Free-Spin-Precession $^{3}$He-$^{129}$Xe Comagnetometer}},
volume = {112},
year = {2014}
}

@book{Bevington1992,
author = {Bevington, Philip R. and Robinson, D. Keith},
booktitle = {McGraw-Hill},
doi = {10.1063/1.4823194},
issn = {08941866},
number = {4},
pages = {415},
publisher = {McGraw-Hill},
title = {{Data Reduction and Error Analysis for the Physical Sciences}},
url = {http://scitation.aip.org/content/aip/journal/cip/7/4/10.1063/1.4823194},
volume = {7},
year = {1992}
}

@article{Beringer2012,
author = {Beringer, J. and Arguin, J. -F. and Barnett, R. M. and Copic, K. and Dahl, O. and Groom, D. E. and Lin, C. -J. and Lys, J. and Murayama, H. and Wohl, C. G. and Yao, W. -M. and Zyla, P. A. and Amsler, C. and Antonelli, M. and Asner, D. M. and Baer, H. and Band, H. R. and Basaglia, T. and Bauer, C. W. and Beatty, J. J. and Belousov, V. I. and Bergren, E. and Bernardi, G. and Bertl, W. and Bethke, S. and Bichsel, H. and Biebel, O. and Blucher, E. and Blusk, S. and Brooijmans, G. and Buchmueller, O. and Cahn, R. N. and Carena, M. and Ceccucci, A. and Chakraborty, D. and Chen, M. -C. and Chivukula, R. S. and Cowan, G. and D'Ambrosio, G. and Damour, T. and de Florian, D. and de Gouv{\^{e}}a, A. and DeGrand, T. and de Jong, P. and Dissertori, G. and Dobrescu, B. and Doser, M. and Drees, M. and Edwards, D. A. and Eidelman, S. and Erler, J. and Ezhela, V. V. and Fetscher, W. and Fields, B. D. and Foster, B. and Gaisser, T. K. and Garren, L. and Gerber, H. -J. and Gerbier, G. and Gherghetta, T. and Golwala, S. and Goodman, M. and Grab, C. and Gritsan, A. V. and Grivaz, J. -F. and Gr{\"{u}}newald, M. and Gurtu, A. and Gutsche, T. and Haber, H. E. and Hagiwara, K. and Hagmann, C. and Hanhart, C. and Hashimoto, S. and Hayes, K. G. and Heffner, M. and Heltsley, B. and Hern{\'{a}}ndez-Rey, J. J. and Hikasa, K. and H{\"{o}}cker, A. and Holder, J. and Holtkamp, A. and Huston, J. and Jackson, J. D. and Johnson, K. F. and Junk, T. and Karlen, D. and Kirkby, D. and Klein, S. R. and Klempt, E. and Kowalewski, R. V. and Krauss, F. and Kreps, M. and Krusche, B. and Kuyanov, Yu. V. and Kwon, Y. and Lahav, O. and Laiho, J. and Langacker, P. and Liddle, A. and Ligeti, Z. and Liss, T. M. and Littenberg, L. and Lugovsky, K. S. and Lugovsky, S. B. and Mannel, T. and Manohar, A. V. and Marciano, W. J. and Martin, A. D. and Masoni, A. and Matthews, J. and Milstead, D. and Miquel, R. and M{\"{o}}nig, K. and Moortgat, F. and Nakamura, K. and Narain, M. and Nason, P. and Navas, S. and Neubert, M. and Nevski, P. and Nir, Y. and Olive, K. A. and Pape, L. and Parsons, J. and Patrignani, C. and Peacock, J. A. and Petcov, S. T. and Piepke, A. and Pomarol, A. and Punzi, G. and Quadt, A. and Raby, S. and Raffelt, G. and Ratcliff, B. N. and Richardson, P. and Roesler, S. and Rolli, S. and Romaniouk, A. and Rosenberg, L. J. and Rosner, J. L. and Sachrajda, C. T. and Sakai, Y. and Salam, G. P. and Sarkar, S. and Sauli, F. and Schneider, O. and Scholberg, K. and Scott, D. and Seligman, W. G. and Shaevitz, M. H. and Sharpe, S. R. and Silari, M. and Sj{\"{o}}strand, T. and Skands, P. and Smith, J. G. and Smoot, G. F. and Spanier, S. and Spieler, H. and Stahl, A. and Stanev, T. and Stone, S. L. and Sumiyoshi, T. and Syphers, M. J. and Takahashi, F. and Tanabashi, M. and Terning, J. and Titov, M. and Tkachenko, N. P. and T{\"{o}}rnqvist, N. A. and Tovey, D. and Valencia, G. and van Bibber, K. and Venanzoni, G. and Vincter, M. G. and Vogel, P. and Vogt, A. and Walkowiak, W. and Walter, C. W. and Ward, D. R. and Watari, T. and Weiglein, G. and Weinberg, E. J. and Wiencke, L. R. and Wolfenstein, L. and Womersley, J. and Woody, C. L. and Workman, R. L. and Yamamoto, A. and Zeller, G. P. and Zenin, O. V. and Zhang, J. and Zhu, R. -Y. and Harper, G. and Lugovsky, V. S. and Schaffner, P.},
doi = {10.1103/PhysRevD.86.010001},
issn = {1550-7998},
journal = {Physical Review D},
month = {jul},
number = {1},
pages = {010001},
title = {{Review of Particle Physics}},
url = {https://link.aps.org/doi/10.1103/PhysRevD.86.010001},
volume = {86},
year = {2012}
}

\end{document}